\documentclass[journal]{IEEEtran}
\newcommand{\subparagraph}{}
\usepackage{titlesec}
\usepackage{subfig}
\usepackage{amsmath}
\usepackage{graphicx}
\usepackage{cite}
\usepackage{mathtools}
\usepackage[usenames, dvipsnames]{color}
\usepackage{algorithm2e}
\usepackage{siunitx}
\usepackage{algpseudocode}
\usepackage{float}
\usepackage{cite}
\usepackage[font=footnotesize]{caption}
\usepackage{tabularx}
\usepackage{array}
\usepackage{blindtext}
\usepackage{soul}
\usepackage{multirow}
\usepackage{balance}
\usepackage{breqn}
\usepackage{soul}
\usepackage[colorinlistoftodos]{todonotes}

\titlespacing{\section}{0pt}{2ex}{1ex}
\titlespacing{\subsection}{0pt}{0.5ex}{0.2ex}
\titlespacing{\subsubsection}{0pt}{0.5ex}{0ex}

\begin{document}

\title{Low Power Unsupervised Anomaly Detection by  Non-Parametric Modeling of Sensor Statistics}
\author{Ahish Shylendra,~\IEEEmembership{Student Member,~IEEE,} Priyesh Shukla,~\IEEEmembership{Student Member,~IEEE,} Saibal Mukhopadhyay,~\IEEEmembership{Fellow,~IEEE,} Swarup Bhunia,~\IEEEmembership{Senior Member,~IEEE,} and Amit~Ranjan~Trivedi,~\IEEEmembership{Member,~IEEE,}}


\maketitle
\begin{abstract}
This work presents AEGIS, a novel mixed-signal framework for real-time anomaly detection by examining sensor stream statistics. AEGIS utilizes Kernel Density Estimation (KDE)-based non-parametric density estimation to generate a real-time statistical model of the sensor data stream. The likelihood estimate of the sensor data point can be obtained based on the generated statistical model to detect outliers. We present CMOS Gilbert Gaussian cell-based design to realize Gaussian kernels for KDE. For outlier detection, the decision boundary is defined in terms of kernel standard deviation ($\sigma_{Kernel}$) and likelihood threshold ($P_{Thres}$). We adopt a sliding window to update the detection model in real-time. We use time-series dataset provided from Yahoo to benchmark the performance of AEGIS. A f1-score higher than 0.87 is achieved by optimizing parameters such as length of the sliding window and decision thresholds which are programmable in AEGIS. Discussed architecture is designed using 45nm technology node and our approach on average consumes $\sim$75 $\mu$W power at a sampling rate of 2 MHz while using ten recent inlier samples for density estimation. \textcolor{red}{Full-version of this research has been published at IEEE TVLSI}

\end{abstract}

\section{Introduction}
\begin{figure}[!t]
     \centering
     \subfloat[][]{\includegraphics[width=.45\linewidth]{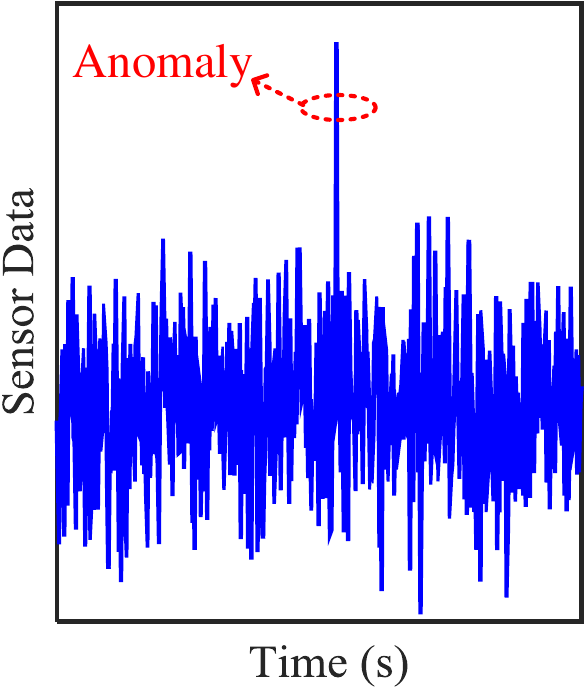}}
     \subfloat[][]{\includegraphics[width=.45\linewidth]{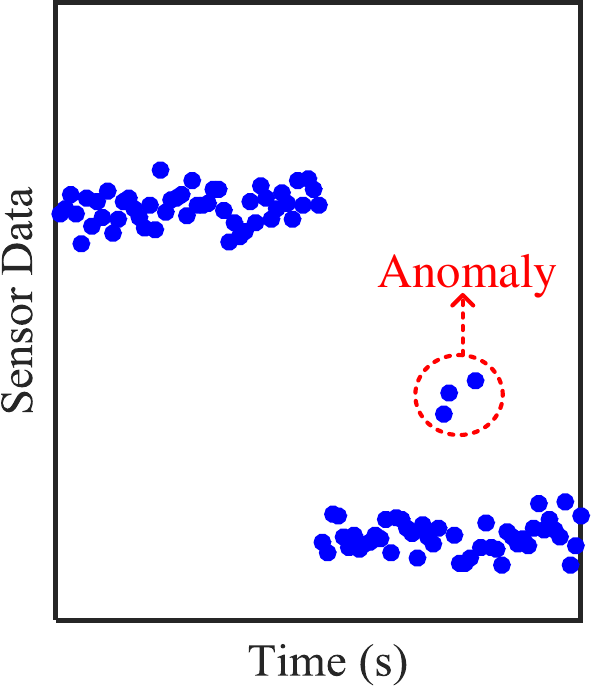}}
     \caption{Illustrations of outliers in streaming sensor data.}
     \label{Anomaly_Illus}
\end{figure}

Sensor networks integrate electronic computation with physical activity for more interactive sensing and control of their application domain. Specifically, a wireless sensor network (WSN) in internet-of-things (IoT) enables more distributed sensing and actuation and thereby enables a heightened awareness and control of their application domain. Sensors in WSNs are also severely constrained in energy, form factor, storage, and computing while operating in unattended and hostile environments. Therefore, sensor measurements are often low quality and are affected by various internal factors, e.g., battery outage and bandwidth limitations, as well as external factors, e.g., environmental adversities and malicious attacks. Meanwhile, a low-quality sensor stream can considerably affect the reliability and accuracy of WSN-based IoT. Therefore, outlier detection and filtering is an important operation for better quality control in IoT \cite{OD_Survey} [Fig. \ref{Anomaly_Illus}]. 

Prior works have utilized various algorithmic schemes such as distance-based, Support Vector Machines (SVM), Bayesian-inference, and Principal Component Analysis (PCA)-based approaches for outlier detection \cite{OD_Survey}. Distance-based methods detect outliers as deviating significantly from the nearest neighbors \cite{OD_Distance}, \cite{OD_Distance_1}. SVM employs model-based outlier detection by fitting a hyperplane to segregate outliers \cite{OD_SVM}. One-class SVM reduces the complexity of a typical SVM to identify outliers as the objects outside the quarter-sphere \cite{OD_1SVM}, \cite{OD_1SVM_1}. Naive Bayesian model detects outliers using probabilistic inference \cite{OD_Bayes}. Bayesian belief networks check for conditional dependencies among observations to detect outliers \cite{OD_Bayes_1}. Although the above algorithmic schemes are successful in outlier detection, they also require complex implementation that typically does not adhere to strict energy, storage, and computing constraints of sensor nodes. Thus, currently, outlier detection is often only performed in remote centralized nodes, which, in turn, incurs transmission energy overhead and affects real-time processing.   

Compared to the above, we present a novel anomaly detection framework, AEGIS, that operates by monitoring the statistics of sensor streams. In AEGIS, each sensor value is characterized against the learned statistics of sensor stream, and outliers are detected as the data values having an extremely low likelihood. AEGIS is real-time since it directly operates on sensor outputs rather than first transforming them to another hyperdimension as in most of the above approaches. AEGIS is low power by using a low complexity mixed-signal design to learn sensor stream statistics. Since AEGIS learns sensor stream statistics non-parametrically, it also applies to data streams of arbitrary statistics. AEGIS updates the sensor's statistical model using a sliding window to grasp temporal variations in the statistics. Prior works \cite{OD_KDE}, \cite{OD_KDE_1},  also utilize a similar approach for outlier detection; however, their evaluation is only algorithmic, whereas we discuss a low power implementation for on-sensor operations. 

AEGIS is also unsupervised and does not require \emph{a priori} knowledge of sensor data distribution. Specifically, we adopt Kernel Density Estimation (KDE) for density synthesis \cite{OD_KDE_2}, \cite{OD_KDE_sina}. We show that kernel cells in KDE-based estimation can be realized using an array of CMOS-based Gilbert Gaussian circuit (GGC). Analog inputs for the GGC are generated using digitally-stored sensor data, a four-bit current steering digital-to-analog converter, and hold cells. A mixed-signal pipeline is discussed to synthesize sensor data density function, which is used to detect outliers in runtime. To benchmark AEGIS, we use time-series dataset from Yahoo \cite{Yahoo}. An outlier detection with an f1-score of more than 0.87 is achieved by optimizing the programmable parameters such as the length of the sliding window and decision thresholds in the discussed implementation. Our approach on average consumes 75$\mu$W for a sampling rate of 2 MHz while using ten recent inlier samples for density estimation.

The rest of the paper is organized as follows: Sec. II provides the background on KDE and its application for outlier detection. Sec. III describes a CMOS GGC design for KDE-based probability density function (PDF) estimation. The overall architecture of AEGIS is described in Sec. IV. Sec. V discusses temperature and process variation-induced inaccuracies in the learned PDF along with optimization of sliding window length and the sampling frequency for reliable PDF estimation. Sec. VI presents the optimization of algorithmic and architectural parameters for efficient outlier detection. Sec. VII discusses adaptability of AEGIS to different IoT applications. Finally, Sec. VIII summarizes the key results and concludes.

\begin{figure}[!t]
     \centering
     \subfloat[][]{\includegraphics[width=.5\linewidth]{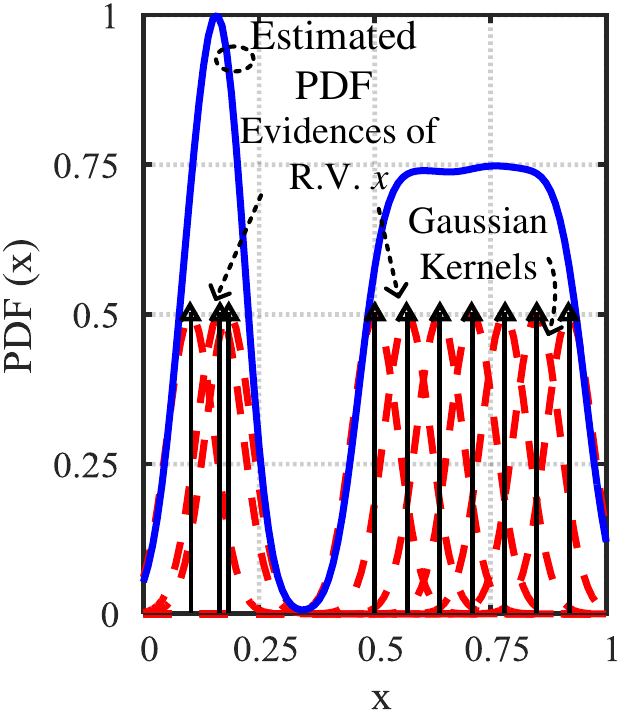}}
     \subfloat[][]{\includegraphics[width=.5\linewidth]{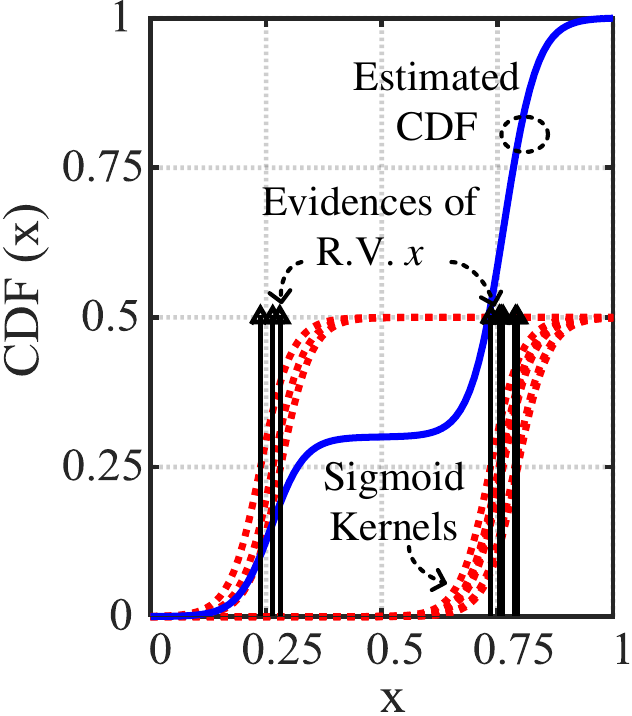}}
     \caption{(a) Non-parametric PDF estimation using Gaussian kernels at observed random variable (RV) samples, (b) Non-parametric CDF estimation using Sigmoid kernels at observed RV samples.}
     \label{KDE_PDFnCDF}
\end{figure}

\section{Background}

\subsection{Kernel density estimation (KDE)}
In KDE, if $x_1$, $x_2$, ...,  $x_N$ are the observed samples of $x$, probability density function (PDF) of $x$, $\tilde{F}(x)$, is estimated as
\begin{equation}
\tilde{F}(x) = \frac{1}{N}\sum_{i=0}^{N-1}k\Big(\frac{x-x_i}{h}\Big)
\end{equation}
Here, $k(x)$ is a kernel function for PDF estimation, $h$ is the kernel function width, and $N$ is the number of observed samples. For probability density function (PDF) estimation, functions such as Gaussian, Uniform, Triangular, and Epanechnikov \cite{Epanechnikov} are applicable as $k(x)$. A cumulative density function (CDF) can also be similarly estimated where functions such as Inverse-tangent and Sigmoid are applicable as $k(x)$ \cite{A1}. Fig. \ref{KDE_PDFnCDF}(a) shows the KDE-based PDF estimation using Gaussian kernels. Fig. \ref{KDE_PDFnCDF}(b) shows CDF estimation using Sigmoid kernels. Parameter $h$ can be optimized to minimize the mean integrated square error (MISE) between the estimated and true PDF/CDF of a random variable (RV). A rule-of-thumb \cite{Density_Est} simplifies the choice of $h$ as $\approx1.06\times\sigma\times N^{-\frac{1}{5}}$ to minimize the MISE, where $\sigma$ is the standard deviation of observed samples of $x$. 

\subsection{Outlier detection by statistical characterization of data}
\begin{figure}[!t]
     \centering
     \includegraphics[width=\linewidth]{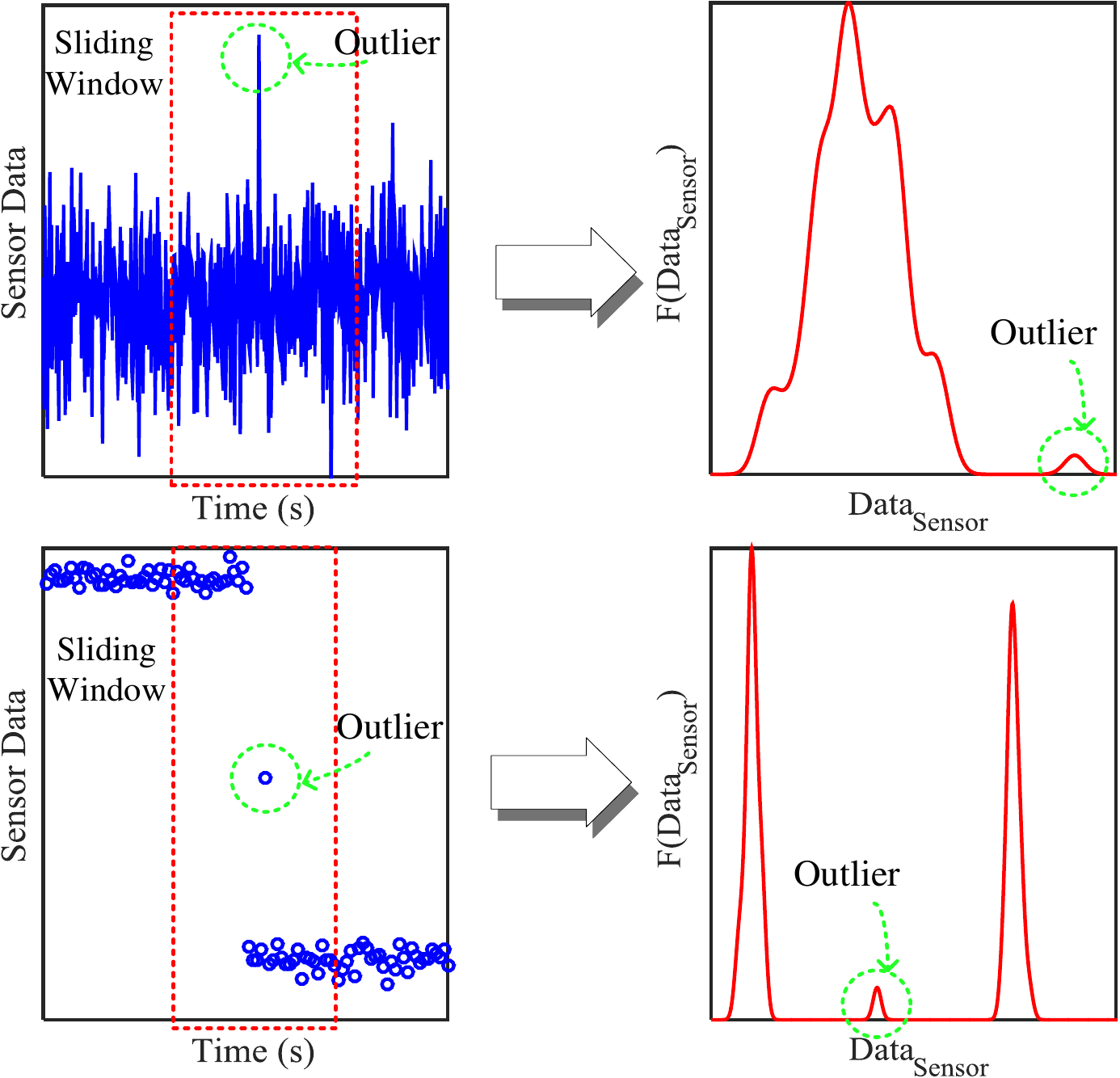}
     \caption{Sliding window for real-time KDE-based density estimation of streaming sensor data considering a limited number of previously validated samples.}
     \label{Sliding_window}
\end{figure}

In Fig. \ref{Sliding_window}, using a sliding window and Eq. (1), the probability density of a subsequent data ($P_{sample}$) can be extracted as \begin{dmath}
P_{sample}= \frac{1}{N}\sum_{i=0}^{N-1}k\Big(\frac{x_{sample}-x_i}{h}\Big)
\end{dmath}
where $x_{sample}$ denotes the incoming sensor data sample and $x_{i}$ are previously validated samples. $x_{sample}$ is classified as an inlier or outlier following the expression below

\begin{figure*}[!ht]
     \centering
     \subfloat[][]{\includegraphics[height=4.8cm]{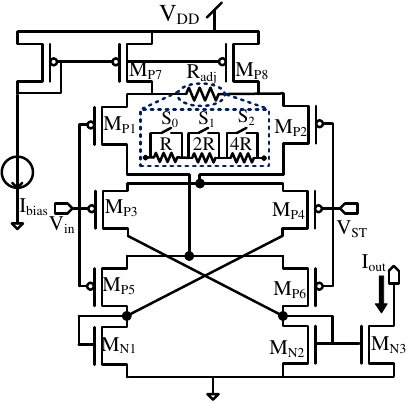}}\quad
     \subfloat[][]{\includegraphics[height=4.8cm]{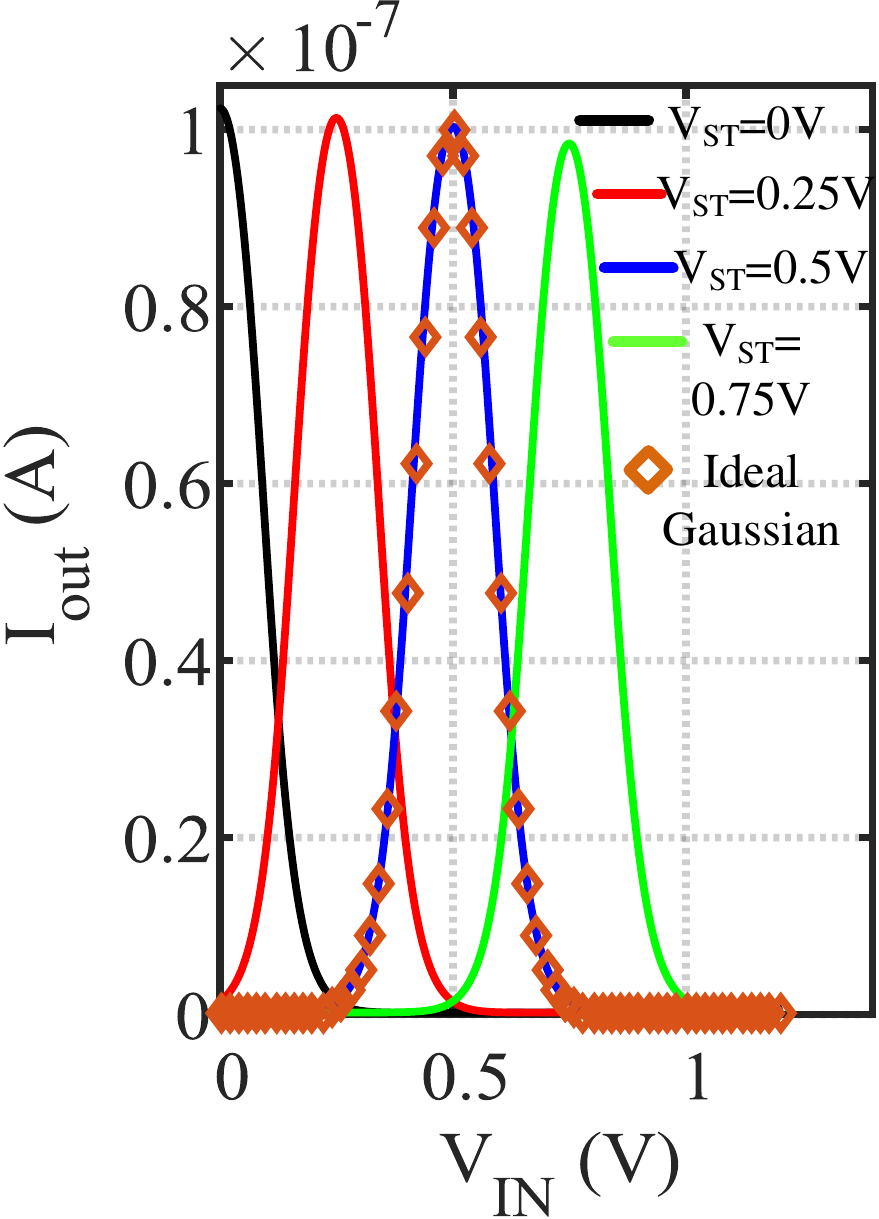}}\quad
     \subfloat[][]{\includegraphics[height=4.8cm]{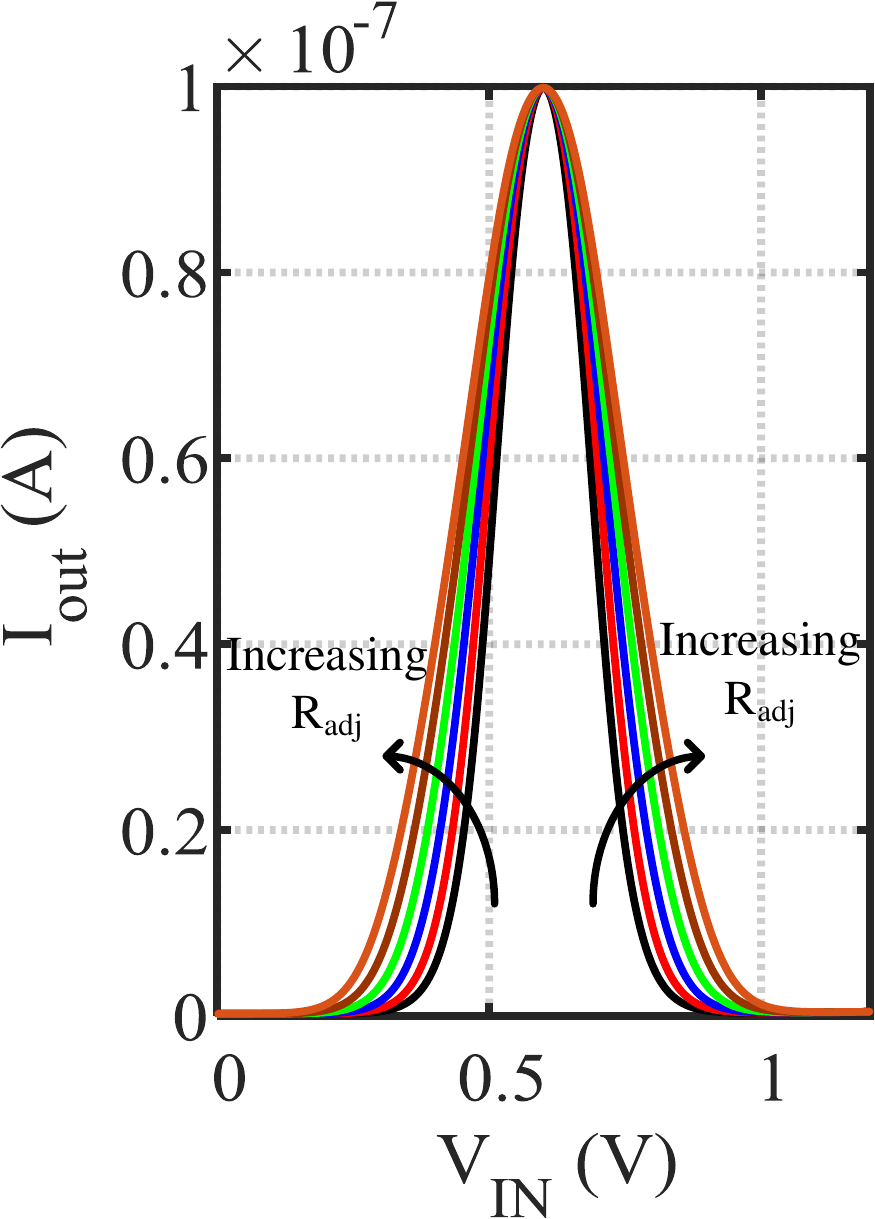}}\quad
     \subfloat[][]{\includegraphics[height=4.8cm]{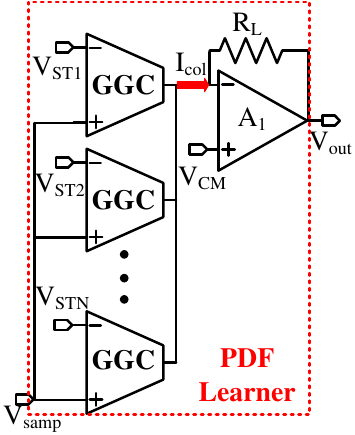}}
     \caption{(a) Gilbert Gaussian circuit to implement kernel cell for PDF estimation (Inset: implementation of $R_{adj}$ using multiple series resistances). (b) Output current of the Gilbert cell follows Gaussian statistics and the Mean ($\mu_{Kernel}$) of the statistics can be programmed using $V_{ST}$. (c) Standard deviation ($\sigma_{kernel}$) of the Gaussian kernel cell can be programmed by varying $R_{adj}$. (d) PDF learner implementation using parallel configuration of GGCs for PDF estimation.}
     \label{OTA}
\end{figure*}

\begin{equation}
  x_{sample}=\left\{ 
   \begin{aligned}
     \text{Outlier}, &&  P_{sample}<P_{Thres} \\
     \text{Inlier}, &&  P_{sample}\geq P_{Thres} 
   \end{aligned}
   \right.
\end{equation}
where $P_{Thres}$ is a probability threshold for outlier detection. For Gaussian kernels, the standard deviation of kernel ($\sigma_{Kernel}$) along with $P_{Thres}$ control the classification boundary. If $x_{sample}$ is an inlier then it is included in the window and the statistical model is updated in a first-in-first-out (FIFO) manner, i.e, by replacing the earliest inlier with $x_{sample}$. Otherwise, $x_{sample}$ is flagged as an anomalous sample. 


\begin{table}[ht]
\centering
\caption{Sizing of the transistors in GGC}
\begin{tabular}{|c|c|c|}
\hline
Transistor & W/L & L\\ \hline
MP1 \& MP2 & 2.4 & L=90nm  \\ \hline
MP3 \& MP4 & 1 & L=90nm  \\ \hline
MP5 \& MP6 & 1 & L=90nm   \\ \hline
MP7 \& MP8 & 0.75 & L=2$\mu$m \\ \hline
MN1 \& MN2 & 1.7 & L=90nm  \\ \hline
MN3        & 3.4 & L=90nm  \\ \hline
\end{tabular}
\end{table}

\section{Gilbert Gaussian Circuit-Based Statistical Density Function Characterization}
\subsection{Gilbert Gaussian Circuit-based Implementation of Kernel Function}
We use a Gilbert Gaussian circuit (GGC) to implement a kernel function for KDE-based PDF synthesis. Henceforth, we use 45nm PDK from \cite{NCSU} for the HSPICE simulations. Fig. \ref{OTA}(a) shows the utilized GGC schematic \cite{SVM_GGC}, \cite{SVM_GGC_1}. Table I provides transistor sizing information for GGC implementation. A test input (equivalent to $x$ in Eq. (1)) is applied at $V_{IN}$ and a sample of $x$ (equivalent to $x_i$ in Eq. (1)) is applied at $V_{ST}$. The test input and $x$ samples in Eq. (1) are mapped to an equivalent voltage in 0$-$3V$_{DD}$/4 range. The mapping range is limited to avoid non-idealities due to input saturation in GGCs while using a simplified and low area/power design. The output current of GGC, I$_{OUT}$, imitates the kernel function output $k(x-x_i/h)$. We achieve 0$-$3V$_{DD}$/4 input range in GGC with PMOS-based input stage.

Fig. \ref{OTA}(b) shows the output current of GGC, $I_{OUT}$, at varying $V_{IN}$ and $V_{ST}$. Since at varying $V_{IN}$, $I_{OUT}$ only has a peak centered around $V_{ST}$, GGC output current is suitable to implement $k(x)$ following Eq. (1). Note that in Fig. \ref{OTA}(b) $I_{OUT}$ of GGC closely follows the Gaussian characteristics, which is an extensively used kernel function for PDF estimation. $R_{adj}$ in Fig. \ref{OTA}(a) enables function width ($h$) programmability of  GGC-implemented kernel. In Fig. \ref{OTA}(c), a higher $R_{adj}$ linearizes transconductance at the input transistors $M_{P1-6}$ resulting in a higher $h$. $R_{adj}$ is implemented as series-connected quantized resistances ($R$, $2R$, and $4R$), where any of the segment resistance can be shorted using the select signals $S_0$--$S_2$ [Inset Fig. \ref{OTA}(a)]. CMOS-based design in \cite{Resistor} can replace the resistors in Fig. \ref{OTA}(a) with transistors, but with added complexity of implementation.

\subsection{GGC-based PDF estimation using kernel density method}
Fig. \ref{OTA}(d) shows PDF learner implementation using parallel configuration of GGCs for estimating PDF following Eq. (1). Each GGC implements a Gaussian kernel function as discussed earlier. The output current, $I_{OUT}$, of all GGCs is summed by shorting them together and the summed current, $I_{COL}$, emulates non-parametric PDF $\widetilde{F}(V_{samp})$. A current-based output in GGCs simplifies summation in Eq. (1) by shorting their output. 
Negative feedback amplifier $A_1$ at the output of parallel GGCs multiplies the column current $I_{COL}$ to $R_L$ producing an output voltage, $V_{OUT}$, proportional to $\widetilde{F}(V_{samp})$ in Eq. (1). Note that $A_1$ also stabilizes the GGC output node potential to $V_{CM}$. OPAMP architecture for $A_1$ is a simple two-stage configuration. OPAMP has been utilized for current to voltage conversion in this architecture, requiring it to have a high slew-rate to quickly respond to variations in input current and to provide stable output. The slew-rate of an OPAMP is limited by its bias current and miller-compensation capacitance between the two stages. Increasing bias current to improve slew-rate degrades energy efficiency. Alternatively, slew-rate can be improved by using a weaker compensation capacitance \cite{Razavi}. 

\section{Online Outlier Detection by Correlating to Data Stream Statistical Density}

\begin{figure}[!ht]
     \centering
     \includegraphics[width=\linewidth]{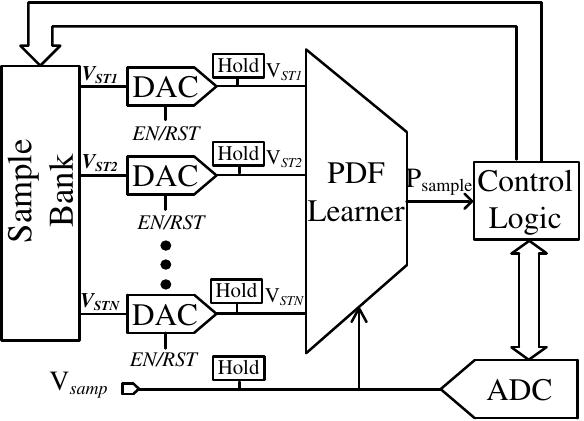}
     \caption{Overall architecture for statistical density estimation and anomaly detection. 
     }
     \label{AEGIS}
\end{figure}

\begin{figure}[!ht]
     \centering
     \includegraphics[width=\linewidth]{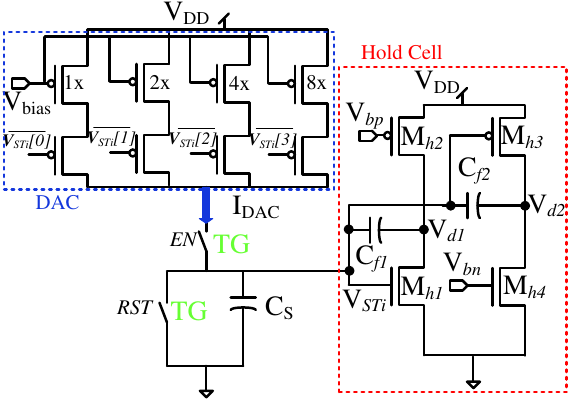}\quad
     \caption{ Integrated 4-bit Current steering DAC \& Hold-cell. [Inset: Transmission Gates are used to control Enable and Reset operations] 
     }
     \label{DAC}
\end{figure}

Fig. \ref{AEGIS} shows the overall architecture of AEGIS comprising of a PDF learner, digital-to-analog converter (DAC) array, analog voltage hold cells, and digital control logic. The design of PDF learner was discussed in Fig. \ref{OTA}. Since PDF learner operates in analog mode, Current-steering DAC (CSDAC) shown in Fig. \ref{DAC} is used to convert the inliers in sample bank (\textbf{\textit{V}}$_{STi}$) ($x_i$ in Eq. (2)) to analog domain. CSDAC uses binary-weighted PMOS current sources to generate a current (I$_{DAC}$) proportional to the digital word of a sample. I$_{DAC}$ charges the sampling capacitance C$_{S}$ on the hold cell [Fig. \ref{DAC}] for a duration T$_{S}$, developing V$_{STi}$. Overdrive voltage constraints on PMOS current sources in CSDAC limit V$_{STi}$ to reliably vary within 0$-$3V$_{DD}$/4 range. Therefore, we use GGC with PMOS input stages for compatibility. Sampling duration T$_{S}$ is controlled using $EN$. Analog samples are retained on the hold cell during PDF learning and outlier detection. Conversely, while updating the detection model, content on the hold cells are reset using $RST$ pulse.

\begin{figure}[ht]
     \centering
     \subfloat[][]{\includegraphics[width=4cm,height=4.5cm]{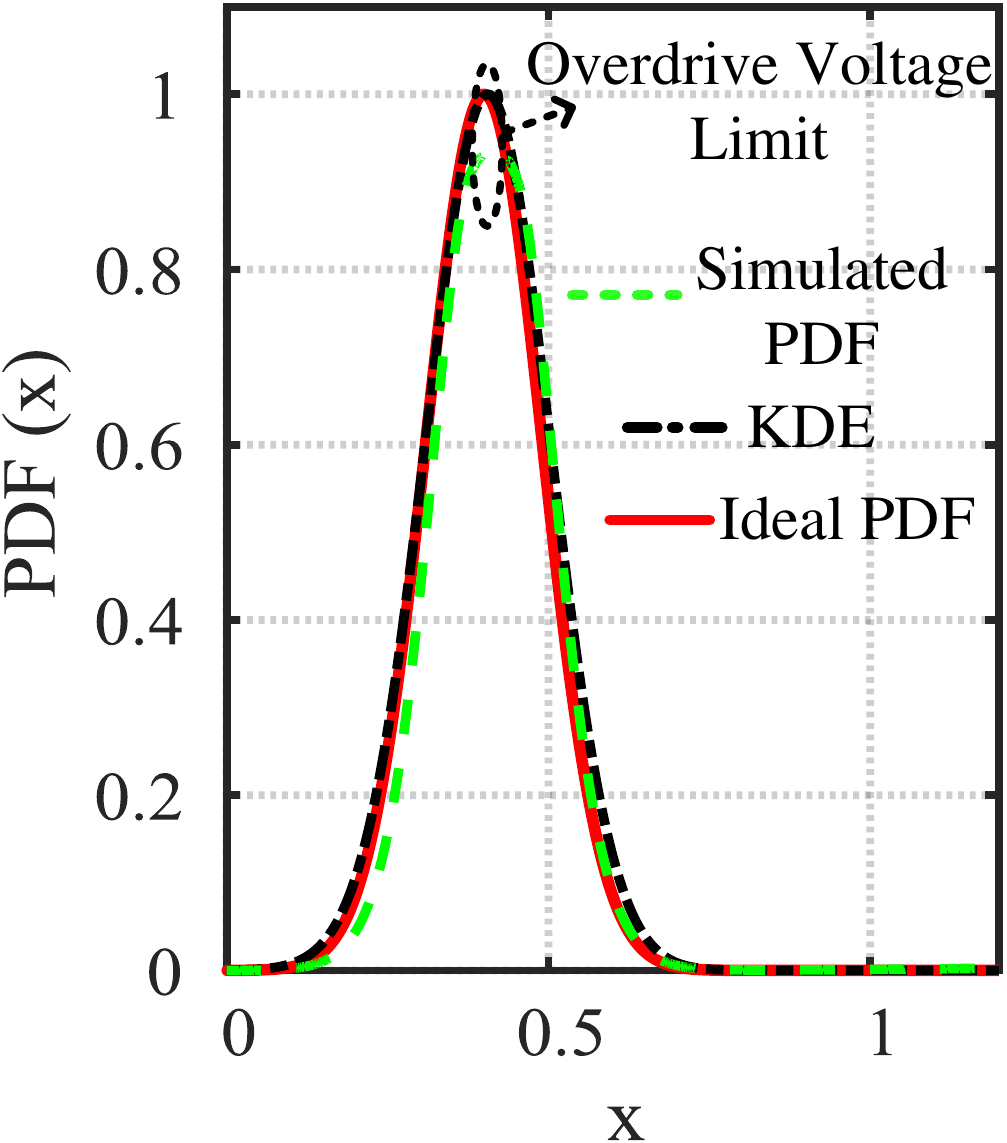}}\quad
     \subfloat[][]{\includegraphics[width=4cm,height=4.5cm]{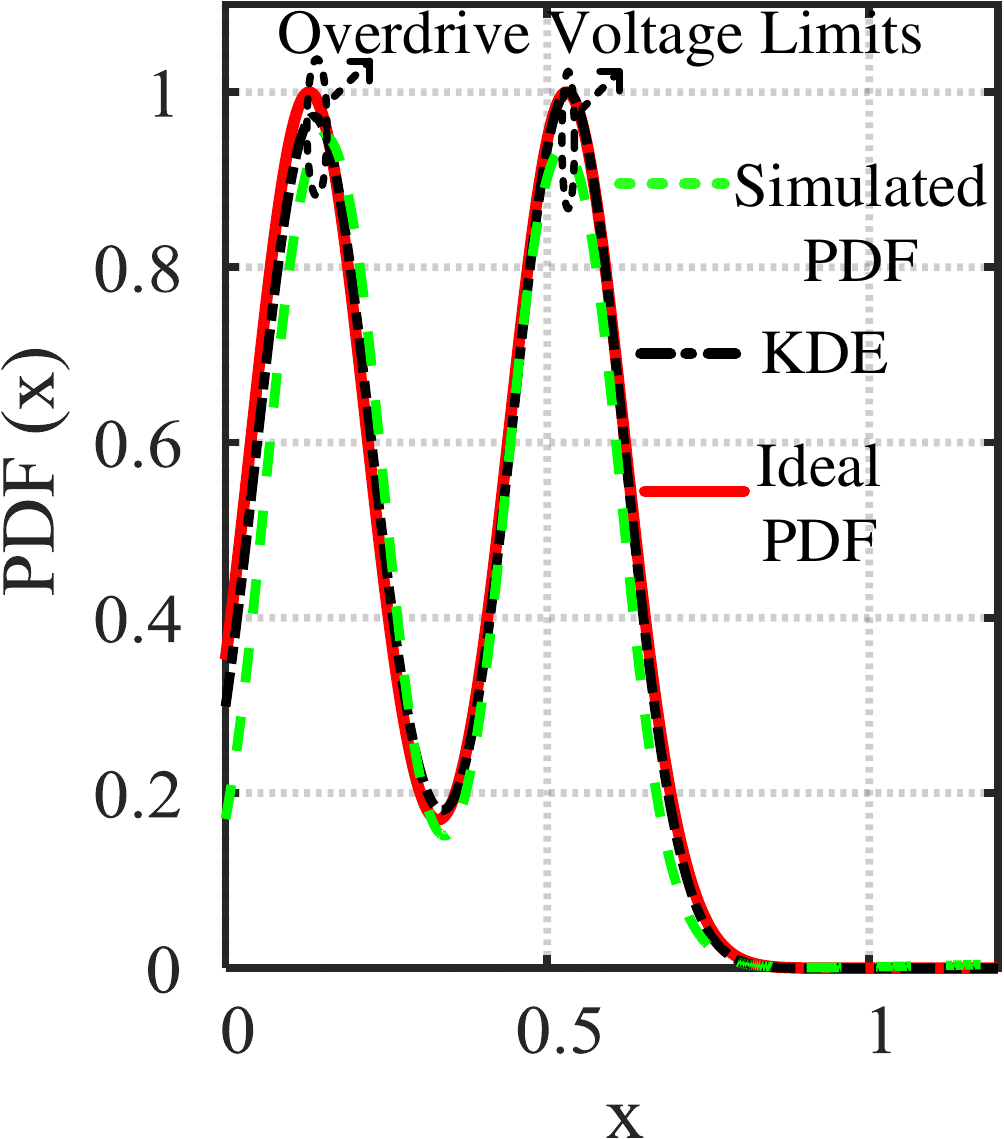}}\quad
     \caption{Non-parametric kernel density estimation enables PDF learning of samples from arbitrary statistics. PDF extracted by PDF learner against ground truth for (a) Gaussian, and (b) Gaussian Mixture distributions.}
     \label{Idealvslearnt}
\end{figure}

Hold cell is shown in Fig. \ref{DAC}. A hold cell consists of common source (CS) amplifiers to improve retention time characteristics. We use both NMOS ($M_{h1}$) and PMOS ($M_{h3}$) input stage based amplifiers to allow 0-3V$_{DD}$/4 range for the hold voltage V$_{STi}$. CS amplifiers are designed to operate in the sub-threshold regime at minimal power consumption. After PDF learning, analog inlier samples held on sampling capacitance (C$_{S}$) will degrade due to thermal leakage in C$_{S}$, C$_{f1}$ \& C$_{f2}$ and sub-threshold leakage in transmission gates. We utilize negative feedback provided by amplifier stages through feedback capacitors (C$_{f1}$ \& C$_{f2}$) to compensate V$_{STi}$. Degrading V$_{STi}$ will alter the bias of $M_{h1}$ and $M_{h3}$, thereby, increases V$_{d1}$ \&  V$_{d2}$ due to negative gain of CS stages. This results in a potential difference across C$_{f1}$ \& C$_{f2}$ which in turn drives a current to restore V$_{STi}$ on C$_{S}$, therefore, improves retention time of the hold cell.


For online outlier detection, AEGIS utilizes a sliding window of past samples. Using a window of $N_{IN}$ latest inliers, PDF learner in Fig. \ref{AEGIS} learns the PDF of the sensor stream. Incoming sample V$_{samp}$ is applied to the input of PDF learner, and PDF learner predicts its likelihood based on the earlier inliers. The control logic in Fig. \ref{AEGIS} determines if the likelihood of V$_{samp}$ is high enough based on the past inliers and if V$_{samp}$ should be accepted as an inlier. If V$_{samp}$ is an inlier, control logic activates ADC and \textbf{\textit{V}}$_{samp}$ (digitized V$_{samp}$) is added to the sample bank where it replaces the first arrived sample. Otherwise, V$_{samp}$ is marked as an outlier.   

\section{Discussions}
In this section, we discuss the accuracy of PDF learner to learn various statistical distributions using a sliding window of past observations. We present an analysis to determine the optimal number of GGCs (N$_{IN}$). Further, the impact of temperature \& process variability and sampling frequency on PDF learner performance is discussed. 
\begin{figure}[!t]
     \centering
     \subfloat[][]{\includegraphics[height=4.8cm]{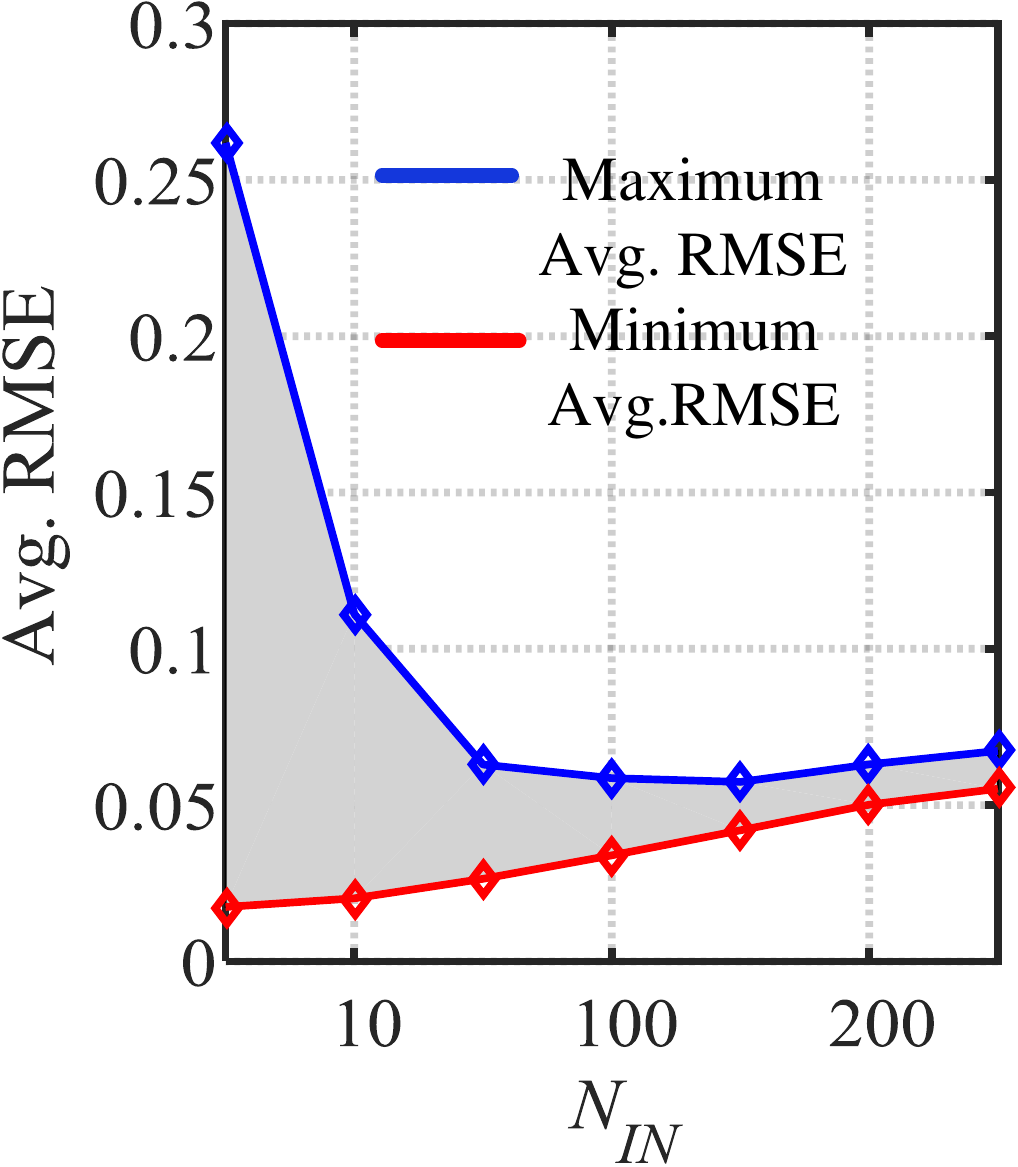}}\quad
     \subfloat[][]{\includegraphics[height=4.8cm]{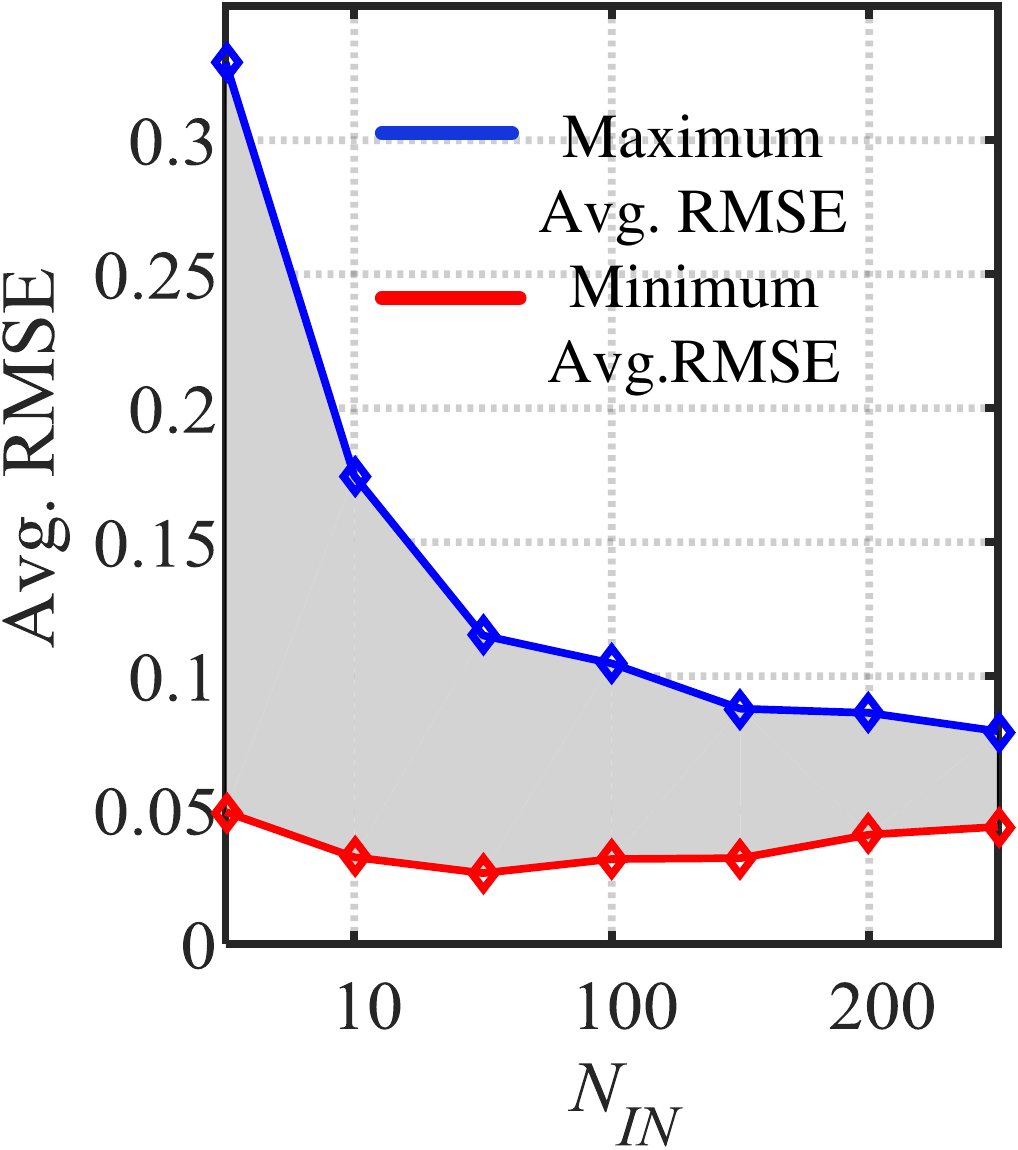}}\quad
     \caption{Avg. RMSE of learned PDF $vs.$ N$_{IN}$ for (a) Gaussian and (b) Gaussian mixture functions.}
     \label{RMSE}
\end{figure}
\subsection{Accuracy to learn various statistical densities}
The accuracy of the PDF learner to learn the statistics of streaming data using a sliding window is analyzed by testing with the sensor data following Gaussian and Gaussian mixture statistics. In Fig. \ref{Idealvslearnt}, ideal PDFs and learned PDFs are compared. The learned PDFs are obtained using Eq. (1) and HSPICE simulations. A hundred randomly drawn samples from the corresponding statistics are used to learn their PDF. 
In Fig. \ref{Idealvslearnt}(a), the samples are drawn from the Gaussian statistics $\mathcal{N}(0.4, 0.05)$. In Fig. \ref{Idealvslearnt}(b), the samples are drawn from the Gaussian Mixture statistics represented by $0.5\times\mathcal{N}(0.15, 0.05)+0.5\times\mathcal{N}(0.55, 0.05)$. In Fig. \ref{Idealvslearnt}, the learned PDFs match with the ground truth except for a small deviation at the peaks due to both the limited samples used for learning and OPAMP overdrive voltage constraints. 

\subsection{Effect of Sliding Window Width}
\begin{figure}[ht]
     \centering
     \subfloat[][]{\includegraphics[height=4.8cm]{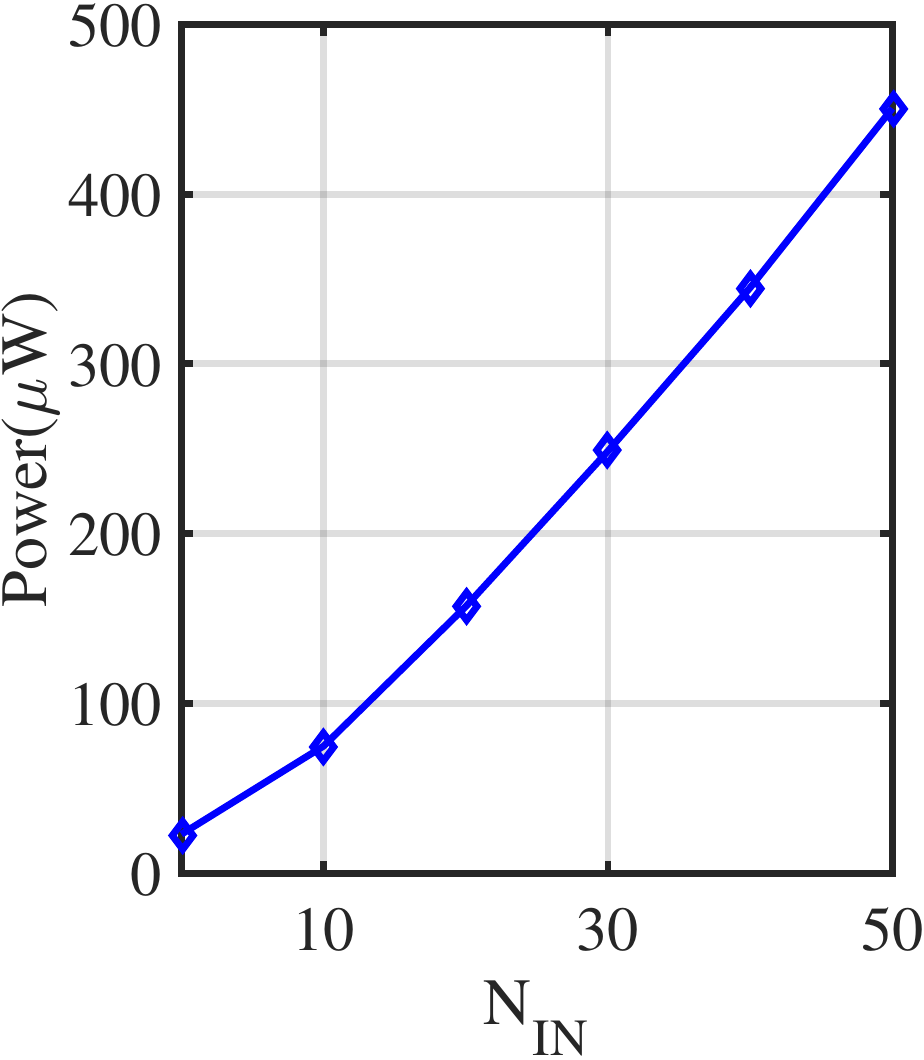}}\quad
     \subfloat[][]{\includegraphics[height=4.8cm]{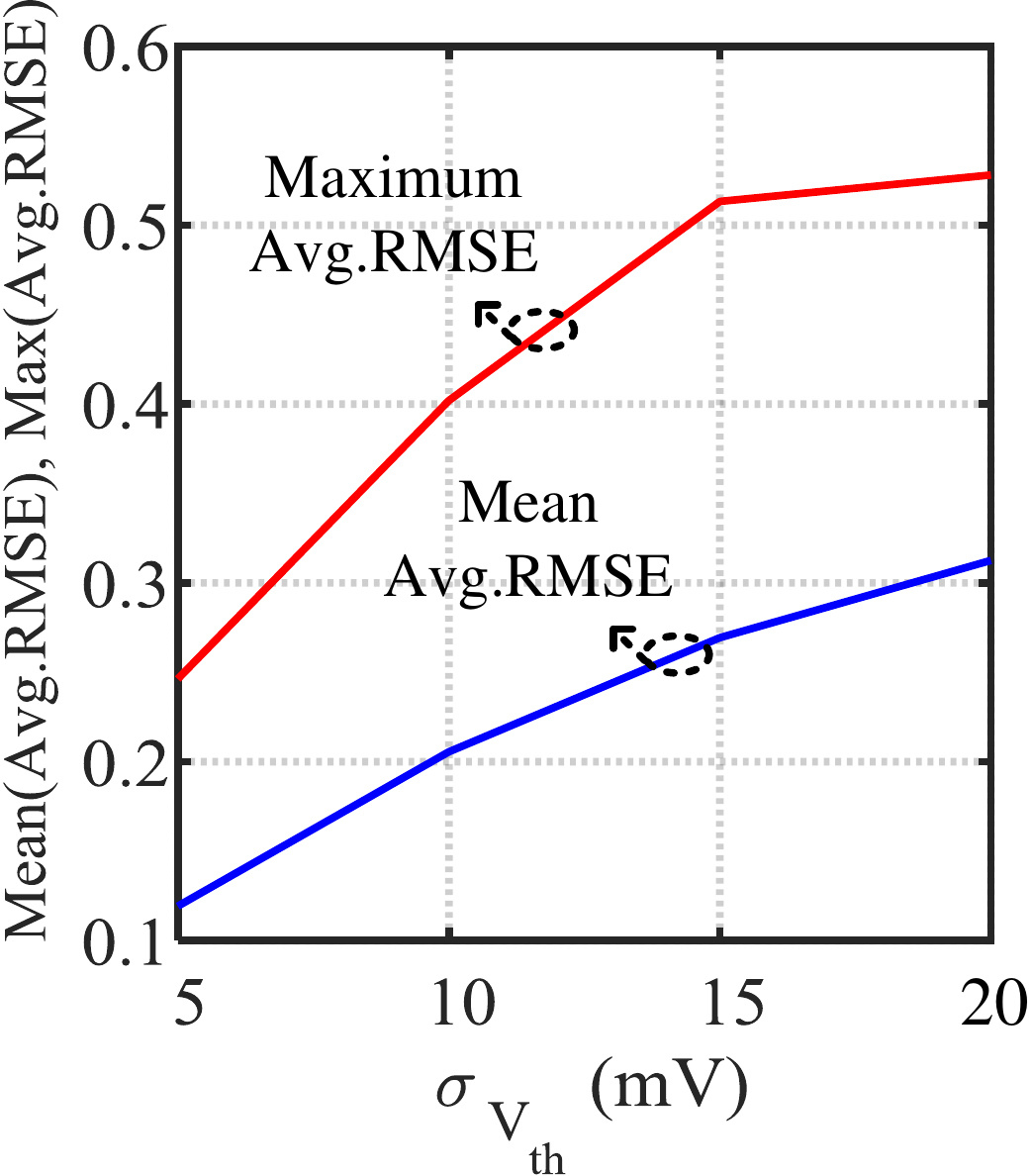}}\quad
     \caption{ (a) Average power consumption of PDF learner and integrated DAC \& hold cell modules for different sliding window length (N$_{IN}$). (b) Impact of process variation on learned PDF is analyzed in terms of Maximum and Mean average RMSE for Gaussian statistics.}
     \label{Process_variation}
\end{figure}

We analyze the robustness of PDF learning in AEGIS considering (i) the number of samples, i.e., N$_{IN}$ and (ii) by limiting the maximum power consumption of the circuit. Using fewer samples is statistically unreasonable for reliable PDF learning. On the other hand, power consumption determines the upper bound on N$_{IN}$. Fig. \ref{RMSE} shows the average root mean square error (Avg.RMSE) between the true and learned PDFs at varying N$_{IN}$ for Gaussian and Gaussian Mixture statistics over a hundred Monte Carlo iterations. The accuracy of the learned PDF improves with increasing N$_{IN}$. However, the power consumption also increases with N$_{IN}$ in Fig. \ref{Process_variation}(a). Therefore, the accuracy-power trade-offs between Fig. \ref{RMSE} \& \ref{Process_variation}(a) determines the optimal N$_{IN}$. 

\subsection{Effect of Process Variability}
To study the impact of process variation on the learned PDF, we consider variations in threshold voltage (V$_{TH}$) of transistors in PDF learner. The resilience of PDF learner against process variation is analyzed in terms of maximum average RMSE (Max$_{Avg.RMSE}$) and mean average RMSE (Mean$_{Avg.RMSE}$) compared against the ground truth. Maximum average RMSE and mean average RMSE are computed for Gaussian statistics over hundred Monte Carlo iterations. In Fig. \ref{Process_variation}(b), V$_{TH}$ variations have a significant impact on the accuracy of learned PDF. Max$_{Avg.RMSE}$ and Mean$_{Avg.RMSE}$ increases with V$_{TH}$ variance. Nonetheless, in Sec. VI, we will discuss that, despite the process variability, a high accuracy outlier detection can be obtained by optimizing the top-level KDE parameters such as $\sigma_{Kernel}$. Process variation induces error in classification hyper-plane, thus, degrades detection efficiency. Optimizing $\sigma_{Kernel}$ can partially restore the classification boundary to suppress the impact of process variability on classification hyper-plane and allows high accuracy detection. 

\begin{figure}[!ht]
     \centering
     \subfloat[][]{\includegraphics[height=4.5cm]{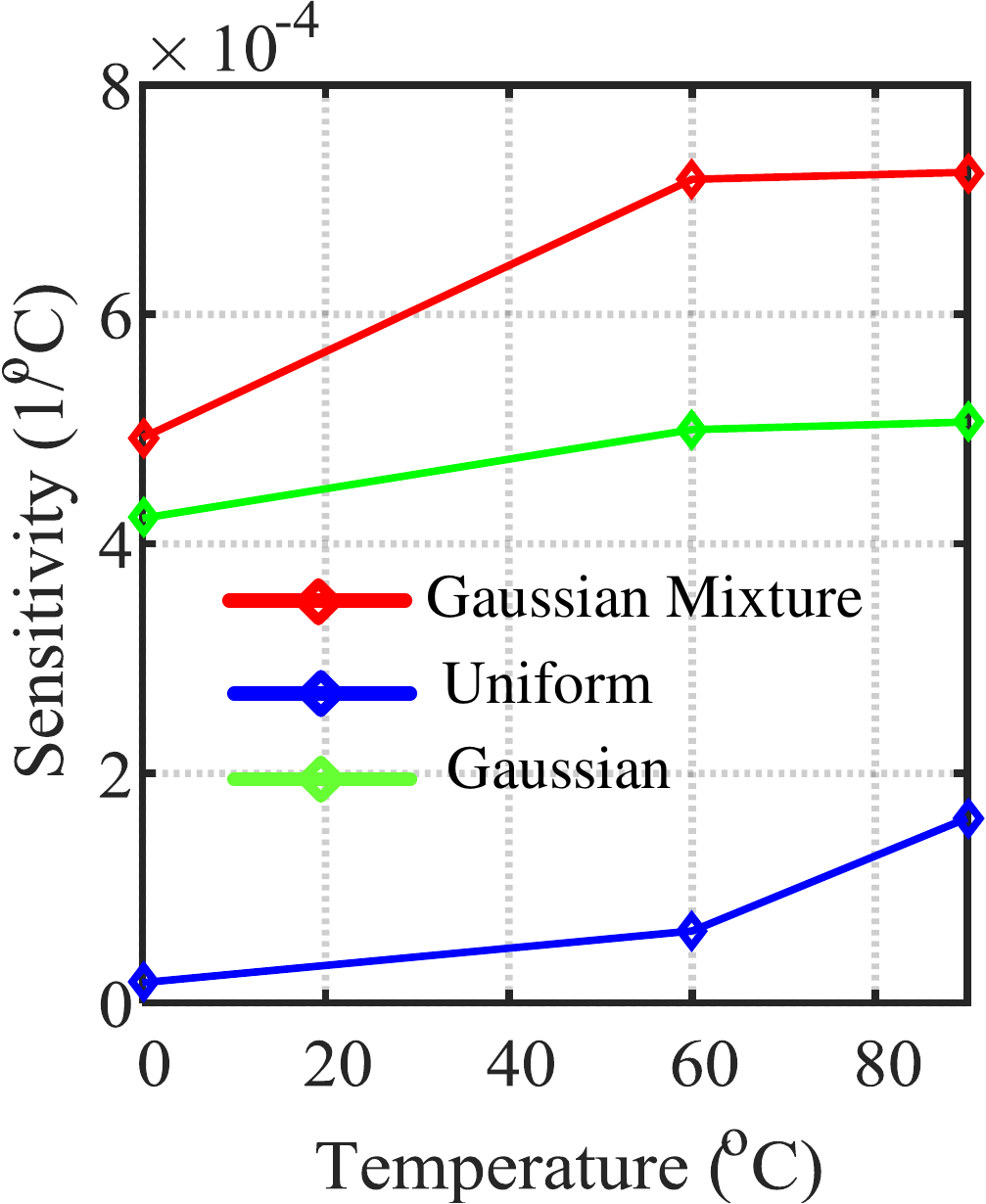}}\quad
     \subfloat[][]{\includegraphics[height=4.5cm]{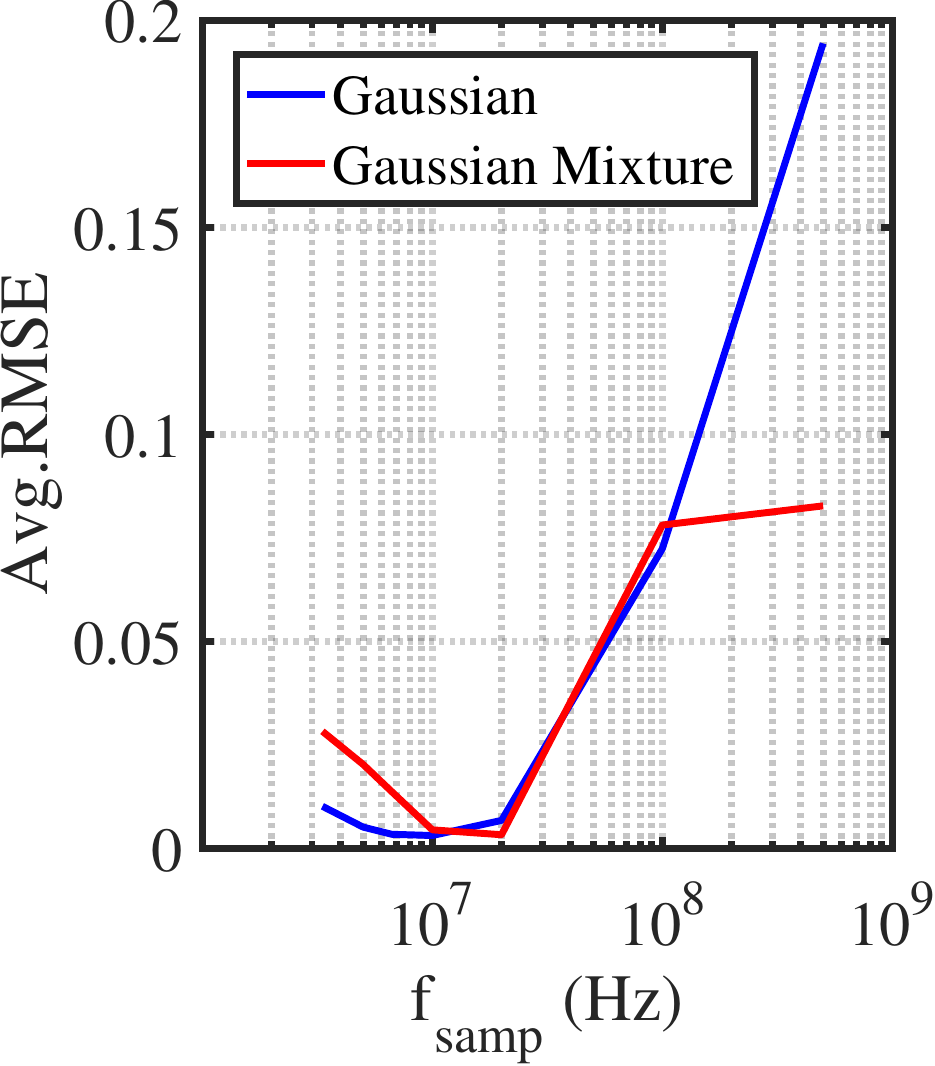}}\quad
     \caption{(a) PDF learner's resilience against temperature variation is analyzed by computing sensitivity of Avg. RMSE with Avg. RMSE at T=30$^o$C as reference. (b) Optimal sampling frequency for reliable operation of PDF learner is determined using average RMSE of learned PDF.}
     \label{Temp_sense}
\end{figure}

\begin{figure*}[!t]
     \centering
     \subfloat[][]{\includegraphics[width=3.3cm, height=4.5cm]{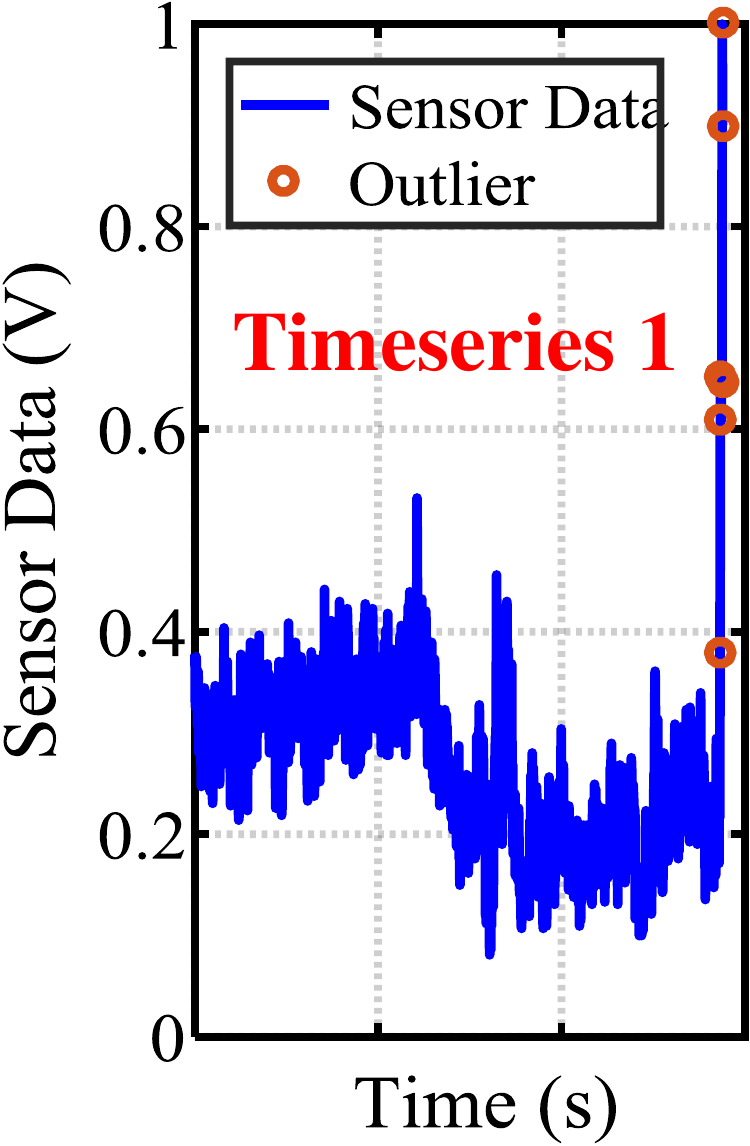}}\quad
     \subfloat[][]{\includegraphics[width=3.3cm, height=4.5cm]{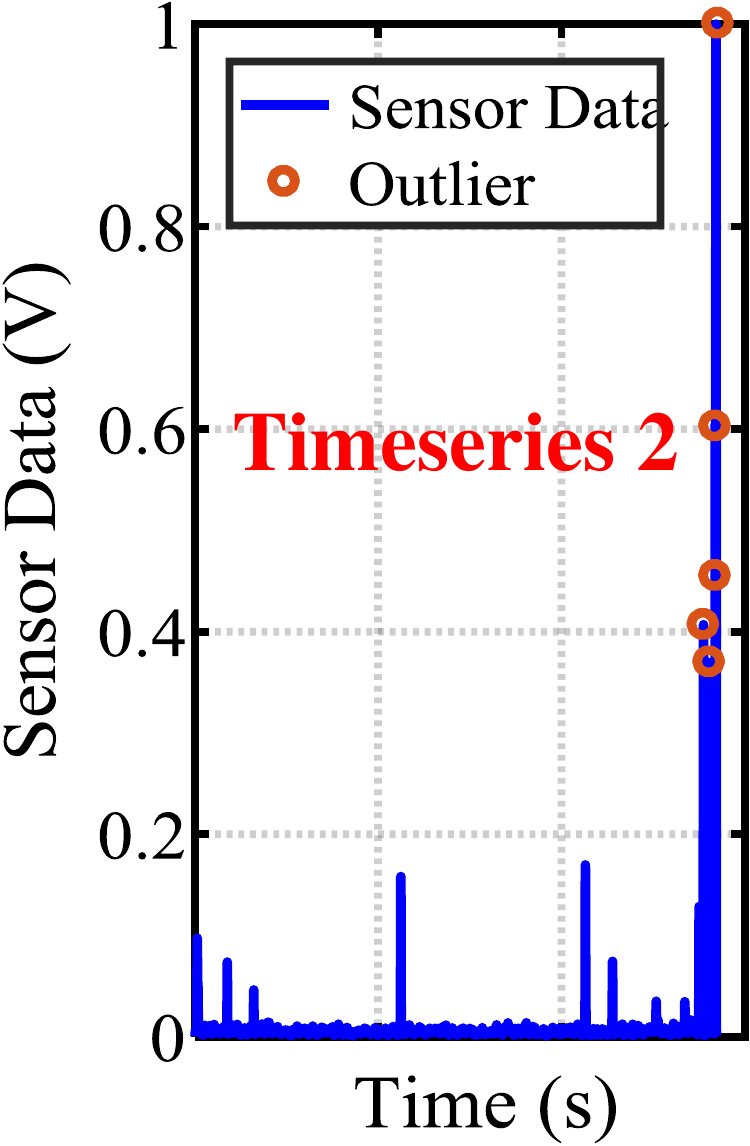}}\quad
     \subfloat[][]{\includegraphics[width=3.3cm, height=4.5cm]{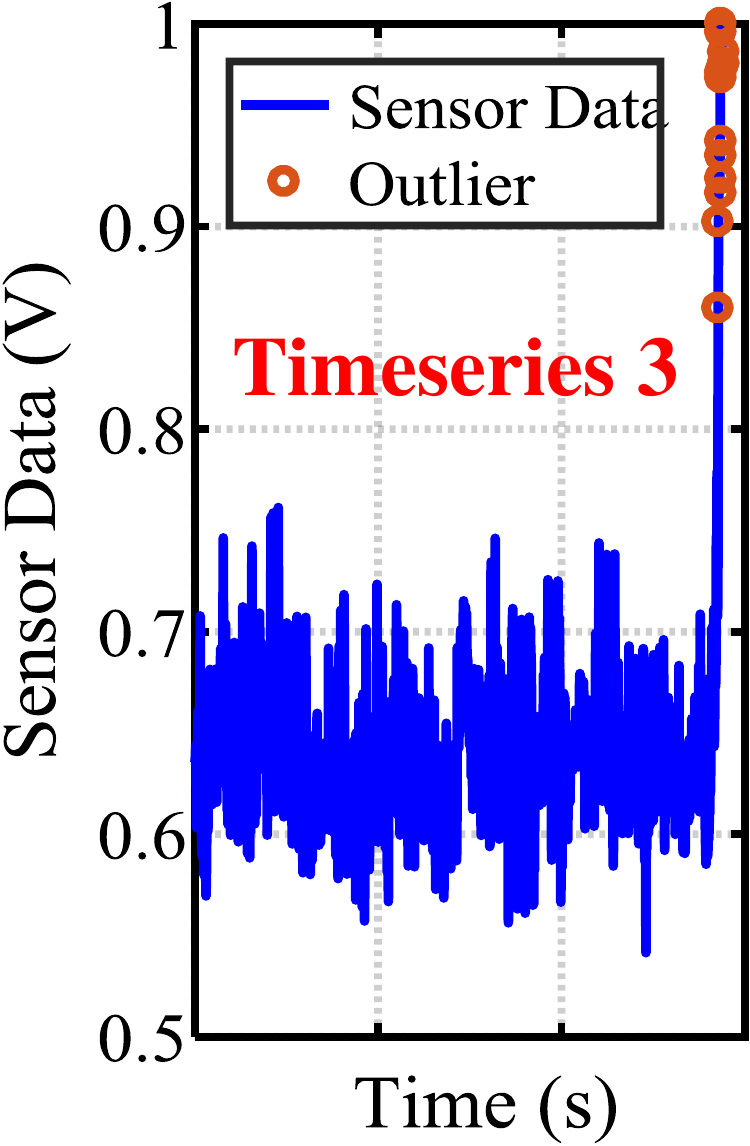}}\quad
     \subfloat[][]{\includegraphics[width=3.3cm, height=4.5cm]{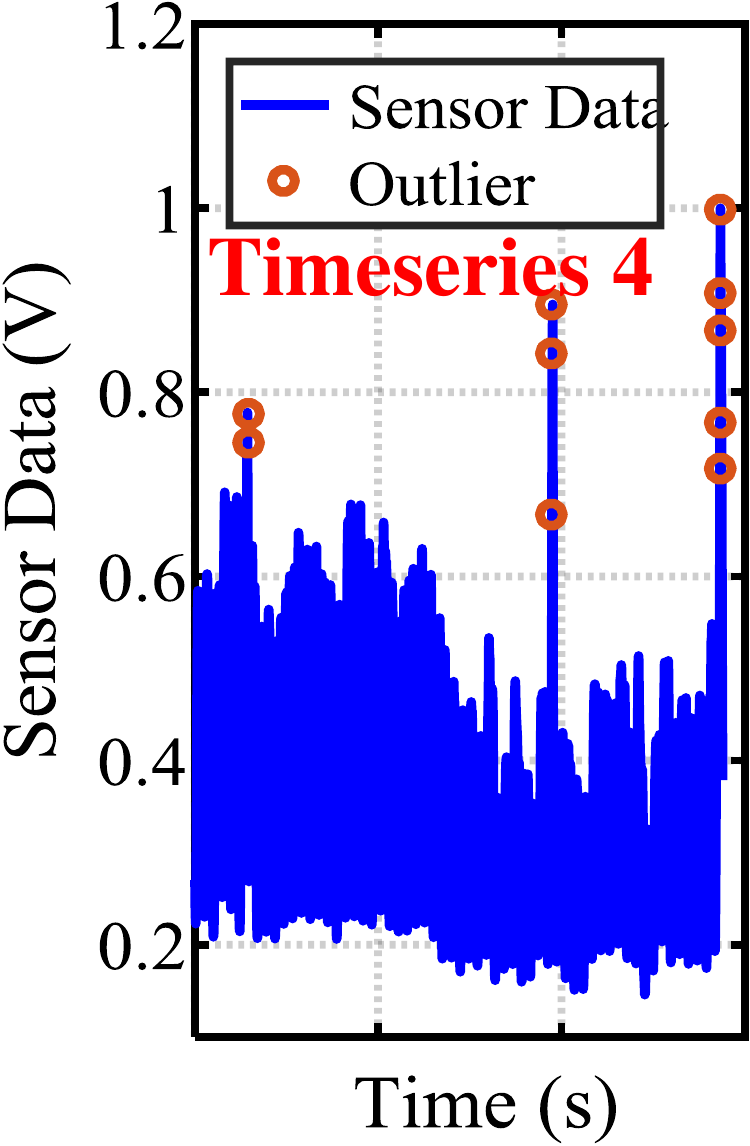}}\quad
     \subfloat[][]{\includegraphics[width=3.3cm, height=4.5cm]{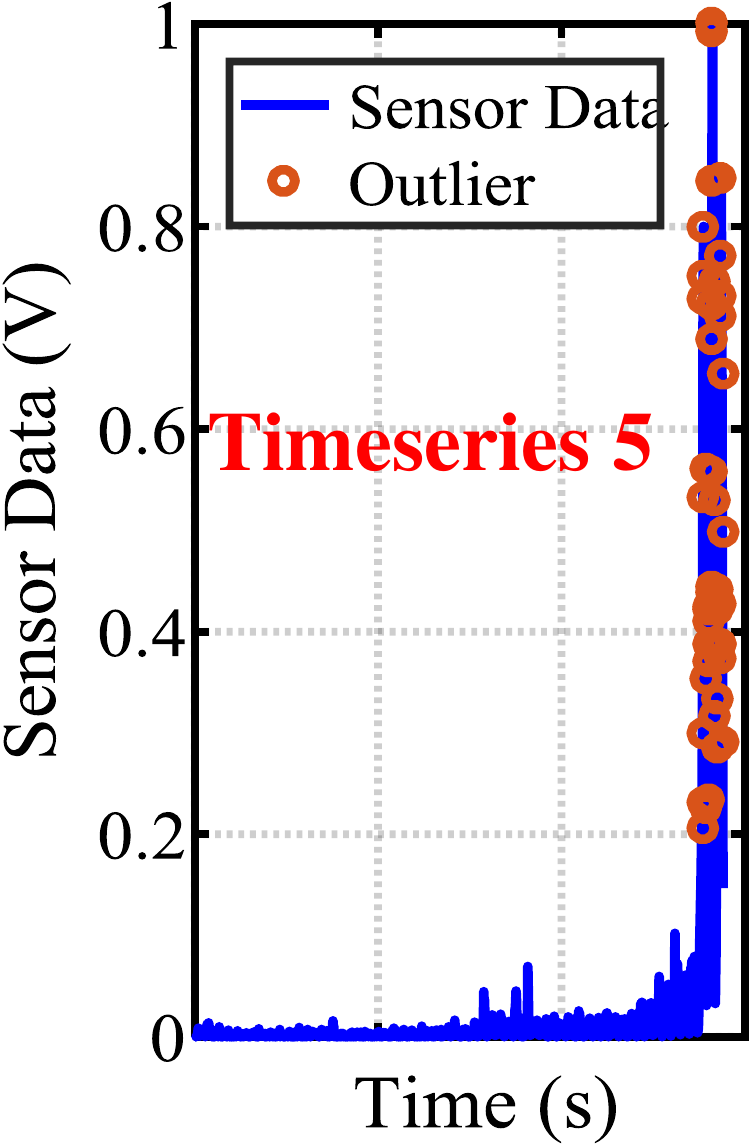}}\quad
     \caption{Different time-series data used to benchmark AEGIS performance.}
     \label{Dataset}
\end{figure*}


\subsection{Effect of Temperature Variability}
WSN edge nodes are often deployed in environments with varying temperatures. In such applications, PDF learner should manifest a high degree of resilience against temperature variations to ensure reliable outlier detection. Impact of temperature variation on the learned PDF is analyzed by computing the sensitivity of average RMSE at varying temperature against room temperature as a reference. Fig. \ref{Temp_sense}(a) shows the temperature sensitivity of learned PDFs for different statistics and is given by sensitivity=$\Delta Avg.RMSE/\Delta T|_{T_{0}=30^oC}$. Temperature variation has a substantial impact on the learned PDF and in Sec. VI, we will show that optimizing KDE parameters enables AEGIS to manifest high resilience against temperature variations. 



\subsection{Effect of Sampling Frequency}
We analyze the impact of sampling frequency on the accuracy of the PDF learner. Ideally, PDF learner should be able to operate at a wide range of frequencies to support varying sampling rates in sensor nodes. However, factors such as the slew-rate of OPAMP and retention time of hold cells impose bounds on the sampling frequency (f$_{samp}$). Fig. \ref{Temp_sense}(b) shows the average RMSE $vs.$ f$_{samp}$ for different sensor stream statistics. While operating at higher frequencies, OPAMP output fails to attain precise values in short duration, resulting in higher average RMSE. On the other hand, samples held on the hold cells degrade over time due to various leakage components, which in turn results in inaccurate PDF estimation at the slower sampling frequency. Therefore, PDF learner functions optimally in a certain f$_{samp}$ range. Various design components in PDF learner, such as hold cells and OPAMP, can be optimized to match f$_{samp}$ to the target sampling rate in a sensor node.

\section{Simulation Results}
In this section, we provide an overview of Yahoo real time-series dataset used to benchmark AEGIS performance. A detailed analysis to determine the optimal parameters for high accuracy outlier detection is presented. We have also investigated the impact of environmental noise on the detection accuracy. Subsequently, we analyze the impact of temperature and process induced variations on the detection efficiency. Furthermore, we compare the power consumption of our approach with comparable digital KDE implementation.

\subsection{Yahoo Real-Time Series Dataset}
We have utilized Yahoo real production traffic time series labeled dataset \cite{Yahoo} to benchmark the performance of the AEGIS. Yahoo dataset consists of both real and synthetic time series data. In synthetic datasets, anomalies are randomly included during synthesis. On the contrary, anomalies in real datasets are identified by analyzing data from the deployed machines. In our experiments, we have benchmarked the performance of our approach using real datasets 4, 6, 10, 15 and 42. These datasets were chosen based on the number of anomalies. We also normalize the datasets to [0, 1] range. Fig. \ref{Dataset} depicts the time-series datasets considered to benchmark AEGIS. We have summarized the characteristics of the datasets in Table II. In Sec. VI(B \& C), Time-series 1 is used to determine optimal parameters. Tables III, IV \& V summarize the performance of the discussed approach with different datasets. 

\begin{table}[!ht]
\centering
\caption{Characteristics of Time series Datasets}
\begin{tabular}{|c|c|c|}
\hline
Time Series & Number of Anomalies & Number of Datapoints \\ \hline
1           & 8                   & 1439                 \\ \hline
2           & 5                   & 1423                 \\ \hline
3           & 13                  & 1439                 \\ \hline
4           & 10                  & 1439                 \\ \hline
5           & 44                  & 1440                 \\ \hline
\end{tabular}
\end{table}

\begin{figure}[!ht]
     \centering
     \subfloat[][]{\includegraphics[width=4cm, height=4.5cm]{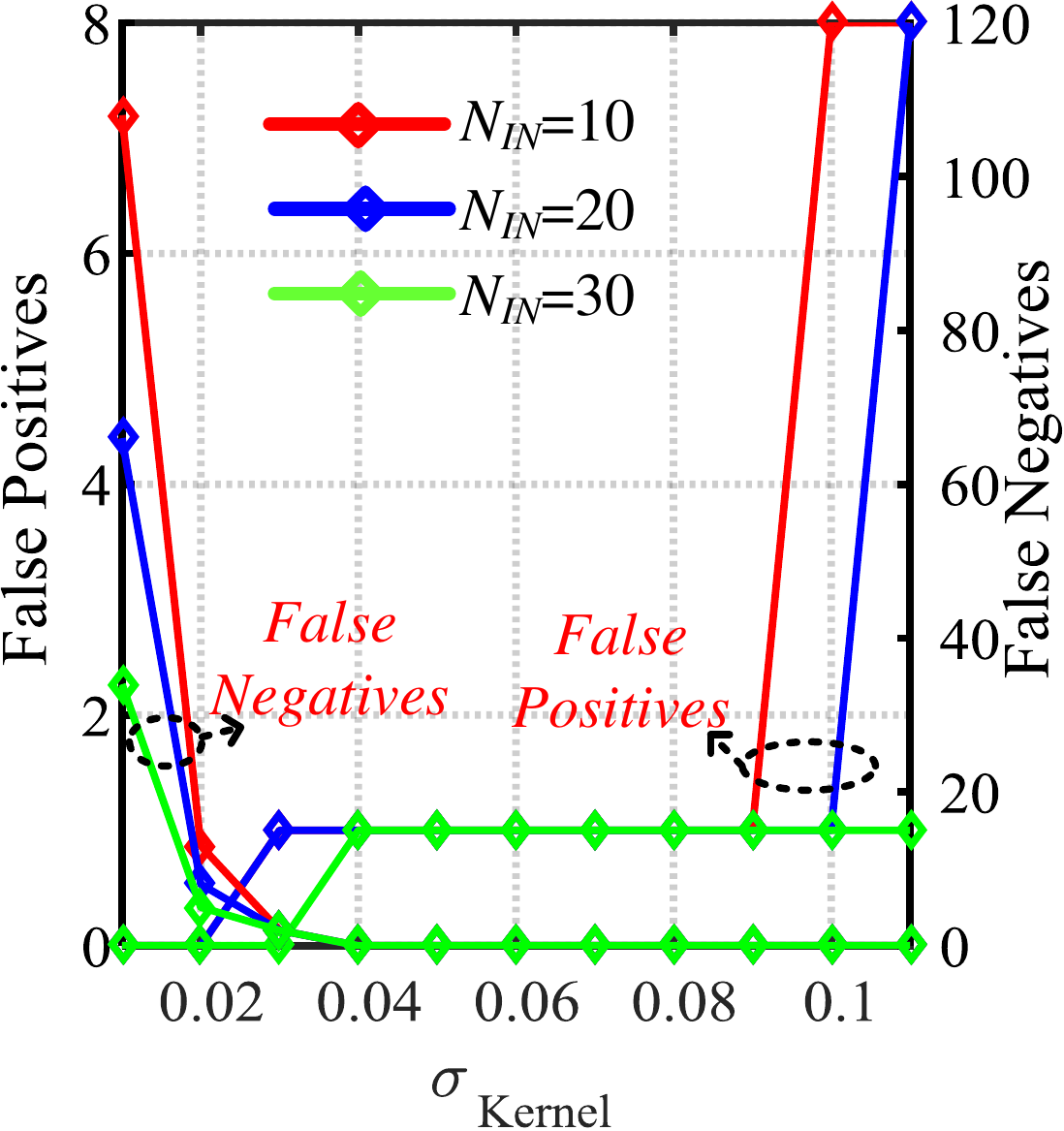}}\quad
     \subfloat[][]{\includegraphics[width=4cm, height=4.5cm]{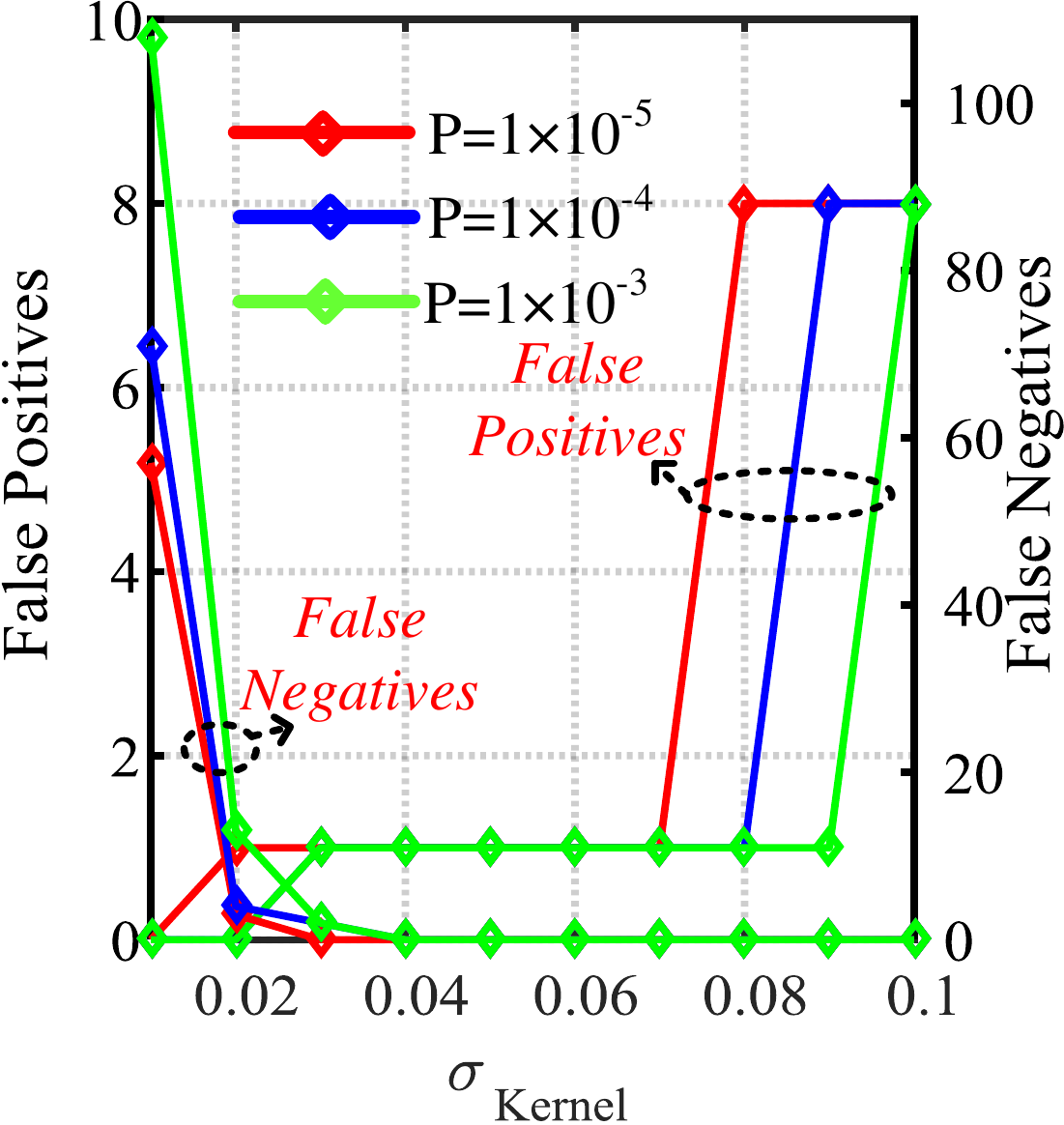}}\quad
     \caption{(a) False positive/false negative performance of AEGIS $vs.$ $\sigma_{Kernel}$ for different sliding window length (N$_{IN}$). (b) Co-optimizing $P_{Thres}$ and $\sigma_{Kernel}$ parameters to minimize F$_{P}$ \& F$_{N}$.}
     \label{Accuarcy_1}
\end{figure}

\subsection{Accuracy Dependence on Algorithmic Parameters}
In this section, we present a study to determine the optimal parameters for KDE to minimize False Positives ($F_{P}$) and False Negatives ($F_{N}$). The examined parameters include the standard deviation of kernels ($\sigma_{Kernel}$), sliding window length (N$_{IN}$), probability threshold ($P_{Thres}$), and DAC resolution. The length of the sliding window (N$_{IN}$) has a significant impact on detection efficiency [Fig. \ref{Accuarcy_1}(a)].
At smaller $\sigma_{Kernel}$, PDF learning using fewer samples results in inaccurate PDF estimation causing inliers to have low-likelihood; resulting in higher $F_{N}$.
At higher $\sigma_{Kernel}$, the tail of the kernels on the neighboring inlier samples overlap. Due to the cumulative effect, the likelihood of all the inlier samples improves [Fig. \ref{Accuarcy_1}(a)]. Therefore, $F_{N}$ decreases at higher $\sigma_{Kernel}$. Nonetheless, $F_{P}$ increases with $\sigma_{Kernel}$ as the likelihood of the outliers also improve due to the cumulative effect. Thus, the optimal $\sigma_{Kernel}$ is needed for efficient detection. Meanwhile, $F_{N}$ and $F_{P}$ performance improves with increasing N$_{IN}$. However, the stringent power budget at the sensor nodes imposes upper-limit on N$_{IN}$ used [Sec. V].

Further, $\sigma_{Kernel}$ along with $P_{Thres}$ defines the hyper-plane segregating inliers and outliers. At higher $P_{Thres}$, using lower $\sigma_{Kernel}$ leads to a tight PDF estimation and degrades likelihood of those inliers closer to the hyper-plane; resulting in higher $F_{N}$. $F_{N}$ performance can be improved by reducing $P_{Thres}$ and/or increasing $\sigma_{Kernel}$ as it alleviates likelihood degradation. On the contrary, higher $\sigma_{Kernel}$ not only improves the likelihood of the inliers but also of the outliers. Also, using lower $P_{Thres}$ results in an inaccurate hyper-plane; increasing $F_{P}$ [Fig. \ref{Accuarcy_1}(b)]. Therefore, trade-off between $F_{N}$ and $F_{P}$ performances determines the optimal $P_{Thres}$ and $\sigma_{Kernel}$.

f1-score metric is used evaluate the anomaly detection performance of AEGIS and is computed as
\begin{equation}
f1-score=\frac{N_A}{N_A+0.5\times(F_N+F_P)}\end{equation}
where N$_A$ denotes the number of anomalies.
Table. III summarizes the f1-score performance for different time-series shown in Fig. \ref{Dataset} with $\sigma_{Kernel}$=0.05 and at varying $P_{Thres}$. Note that a fixed $P_{Thres}$ is not optimal for all test-cases and should be selected based on high-level characteristics of sensor stream. 


\begin{table}[ht]
\centering
\caption{f1-score performance of the discussed architecture for different time-series data with $\sigma_{Kernel}$=0.05.}
\scalebox{0.75}{
\begin{tabular}{|c|c|c|c|c|c|}
\hline
\multirow{2}{*}{Time Series} & \multicolumn{5}{c|}{f1-score}                                                \\ \cline{2-6} 
                             & P$_{Thres}$=10$\mu$ & P$_{Thres}$=100$\mu$ & P$_{Thres}$=1m & P$_{Thres}$=10m & P$_{Thres}$=100m \\ \hline
1                            & \textbf{\textcolor{black}{0.9412}}            & \textbf{\textcolor{black}{0.9412}}               & \textbf{\textcolor{black}{0.9412}}         & 0.8421        & 0.8         \\ \hline
2                            & 0.9091            & 0.9091             & \textbf{\textcolor{black}{1}}           & 0.83        & 0.67         \\ \hline
3                            & 0.67            & 0.67             & 0.67       & 0.67        & \textbf{\textcolor{black}{1}}             \\ \hline
4                            & 0.9524            & 0.9524            & \textbf{\textcolor{black}{1}}        & 0.74       & 0.0485         \\ \hline
5                            & 0.69           & 0.72             & 0.73      & 0.76       & \textbf{\textcolor{black}{0.95}}            \\ \hline
\end{tabular}}
\end{table}

\begin{figure}[!t]
     \centering
     \subfloat[][]{\includegraphics[width=4cm, height=4.5cm]{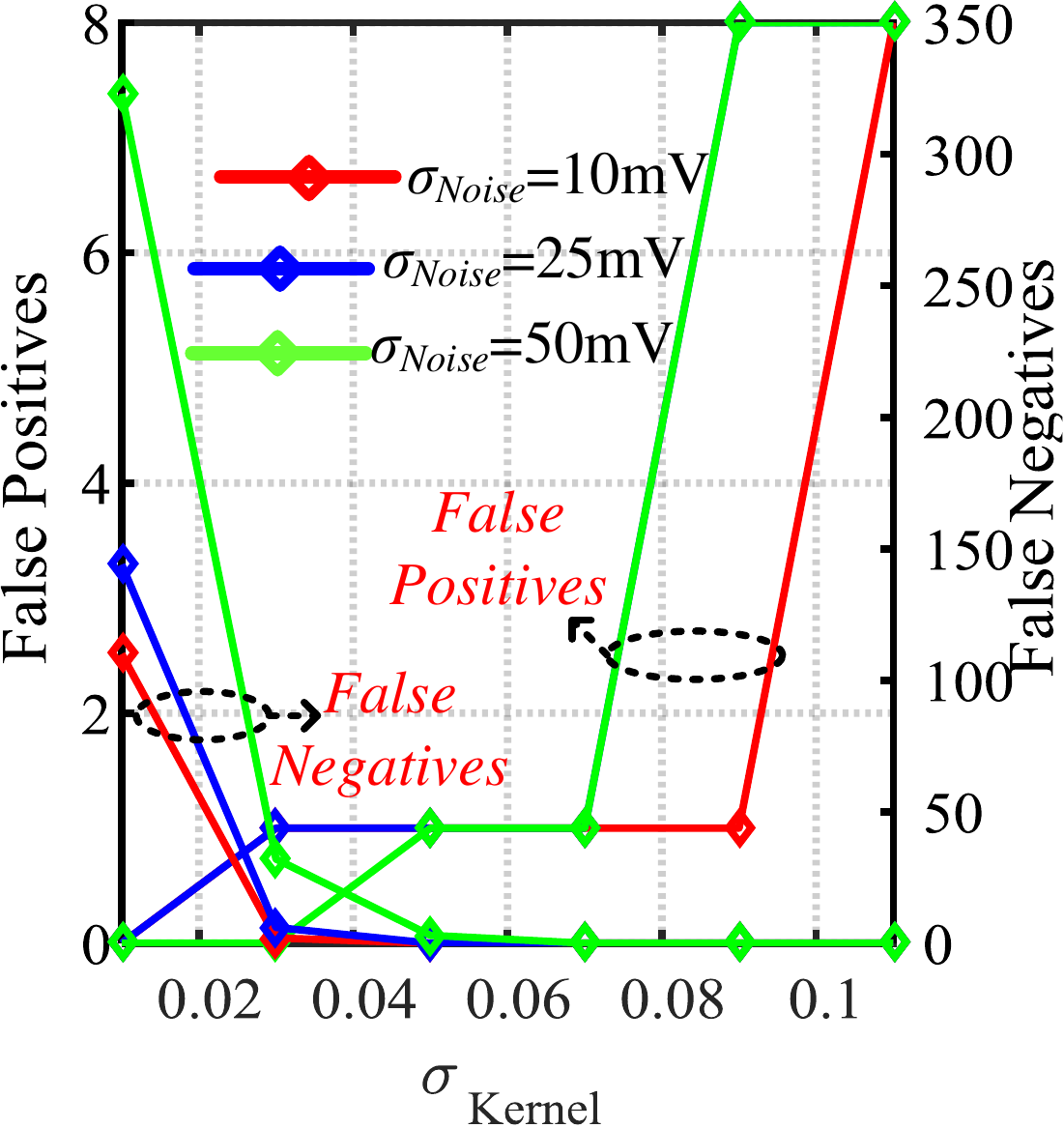}}\quad
      \subfloat[][]{\includegraphics[width=4cm, height=4.5cm]{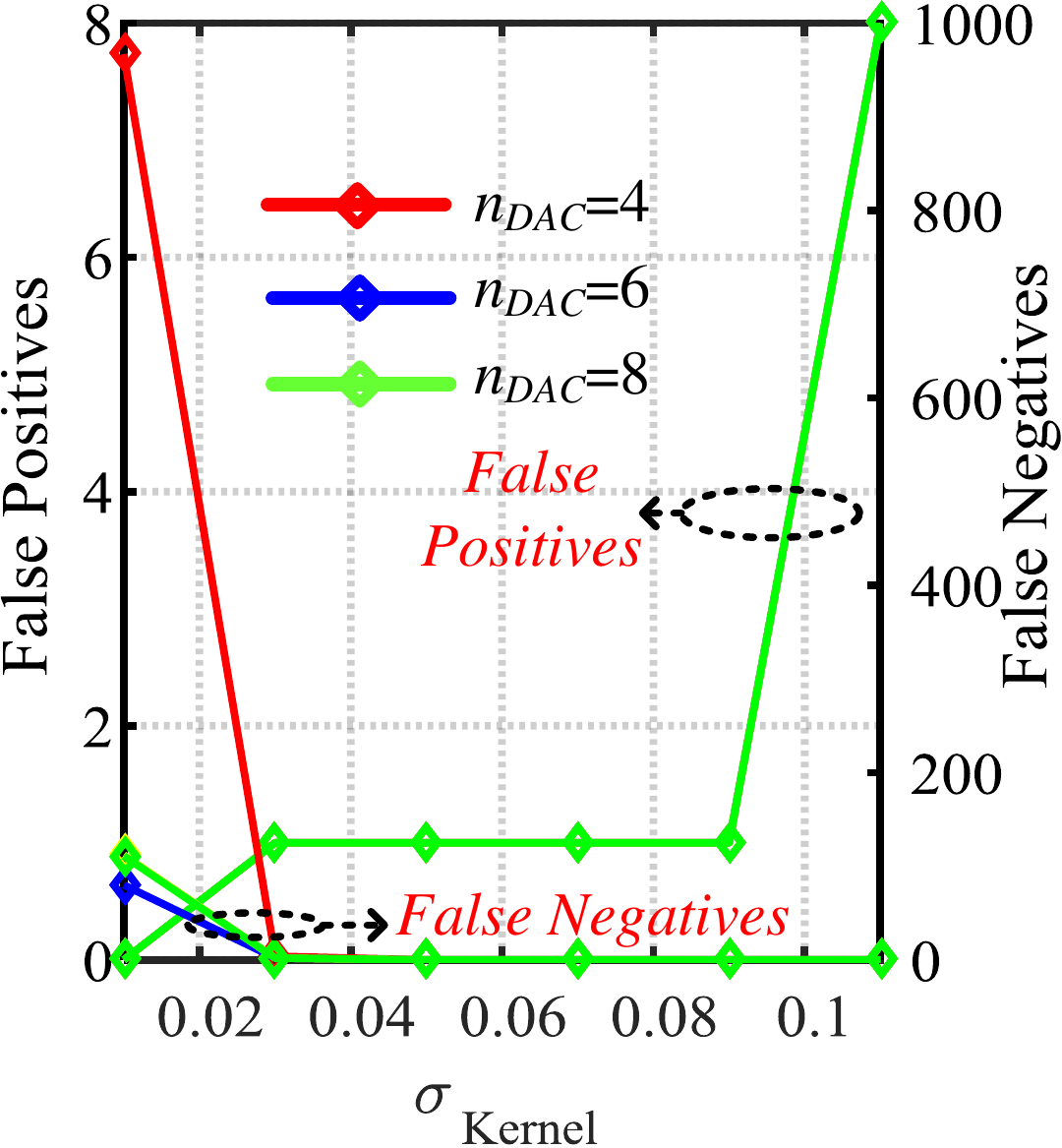}}\quad
     \caption{ (a) Environmental noise significantly affects AEGIS performance. Programming $\sigma_{Kernel}$ allows to improve AEGIS's resilience against noise and minimize false inference, and (b) Optimizing DAC resolution and $\sigma_{Kernel}$ to minimize false inference rate as well as DAC power consumption.}
     \label{Accuracy_2}
\end{figure}

\subsection{Accuracy Dependence on Implementation Parameters}

To study the impact of noise on detection efficiency, we have considered additive Gaussian noise. Fig. \ref{Accuracy_2}(a) shows $F_{N}$ and $F_{P}$ performances for different Gaussian noise variance ($\sigma_{noise}$).
False negatives/positives increases with the noise level; however, by programming $\sigma_{Kernel}$, we can achieve high detection efficiency even in extremely noisy conditions. 

Next, we have studied the impact of DAC resolution on detection efficiency. In Fig. \ref{Accuracy_2}(b), utilizing fewer bits to represent streaming data leads to significant loss of information due to higher quantization error; degrading the likelihood of the inlier samples near the hyper-plane which in turn leads to higher $F_{N}$.
Similarly, outliers closer to the hyper-plane will be inaccurately assigned higher likelihood due to the quantization error; increasing $F_{P}$. It should be noted that quantization error at moderate DAC resolution affects only inliers and outliers closer to the hyper-plane. Therefore, inliers and outliers away from the hyper-plane will be affected only by KDE parameters.   
On the other hand, higher DAC resolution degrades the energy efficiency of the architecture as the DAC power consumption increases exponentially with the resolution. Optimally, we have considered a DAC with a 4-bit resolution to minimize power consumption while retaining higher detection efficiency by programming $\sigma_{Kernel}$. 

\begin{figure}[!t]
     \centering
\subfloat[][]{\includegraphics[width=4cm, height=4.5cm]{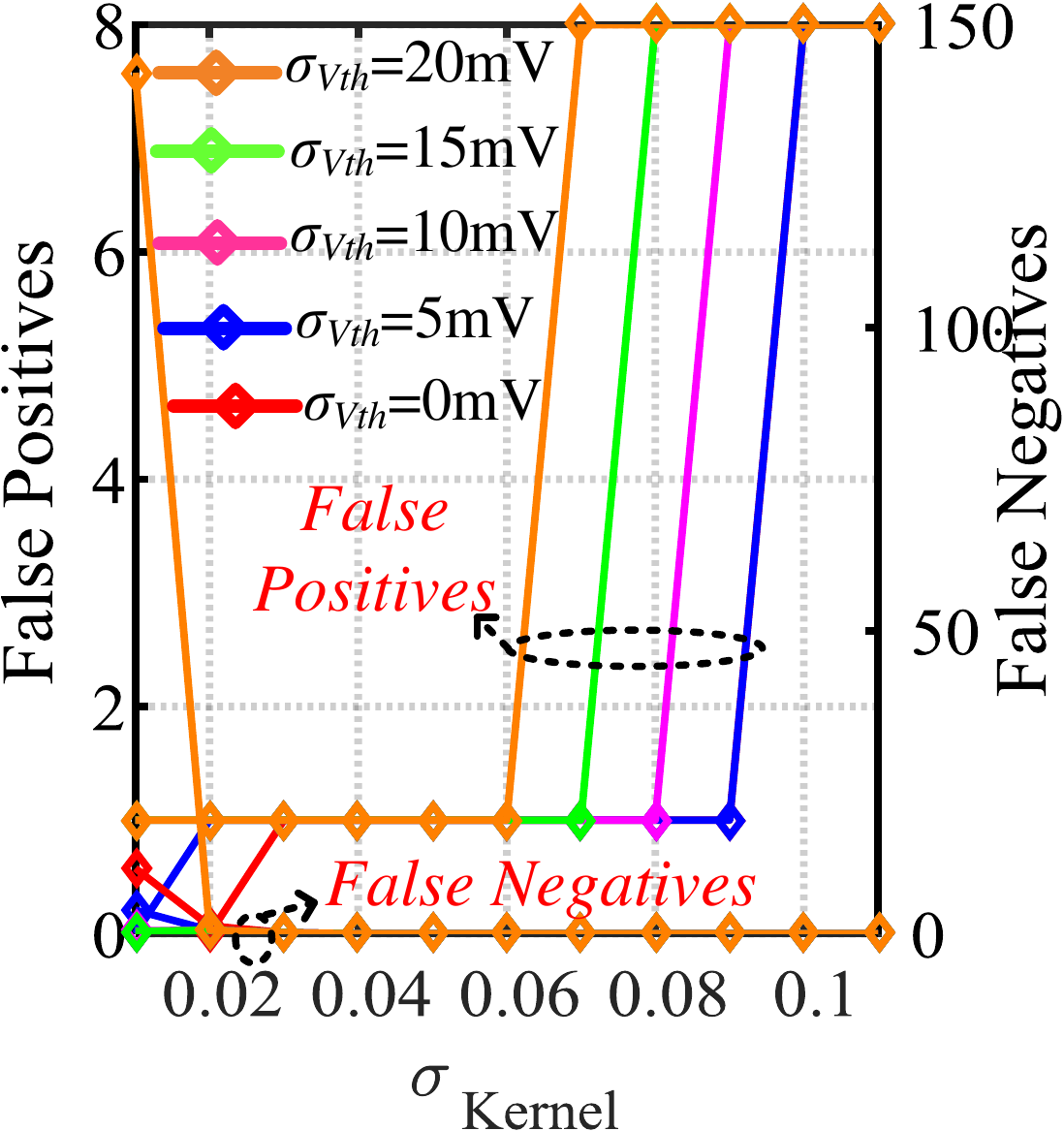}}\quad
\subfloat[][]{\includegraphics[width=4cm, height=4.5cm]{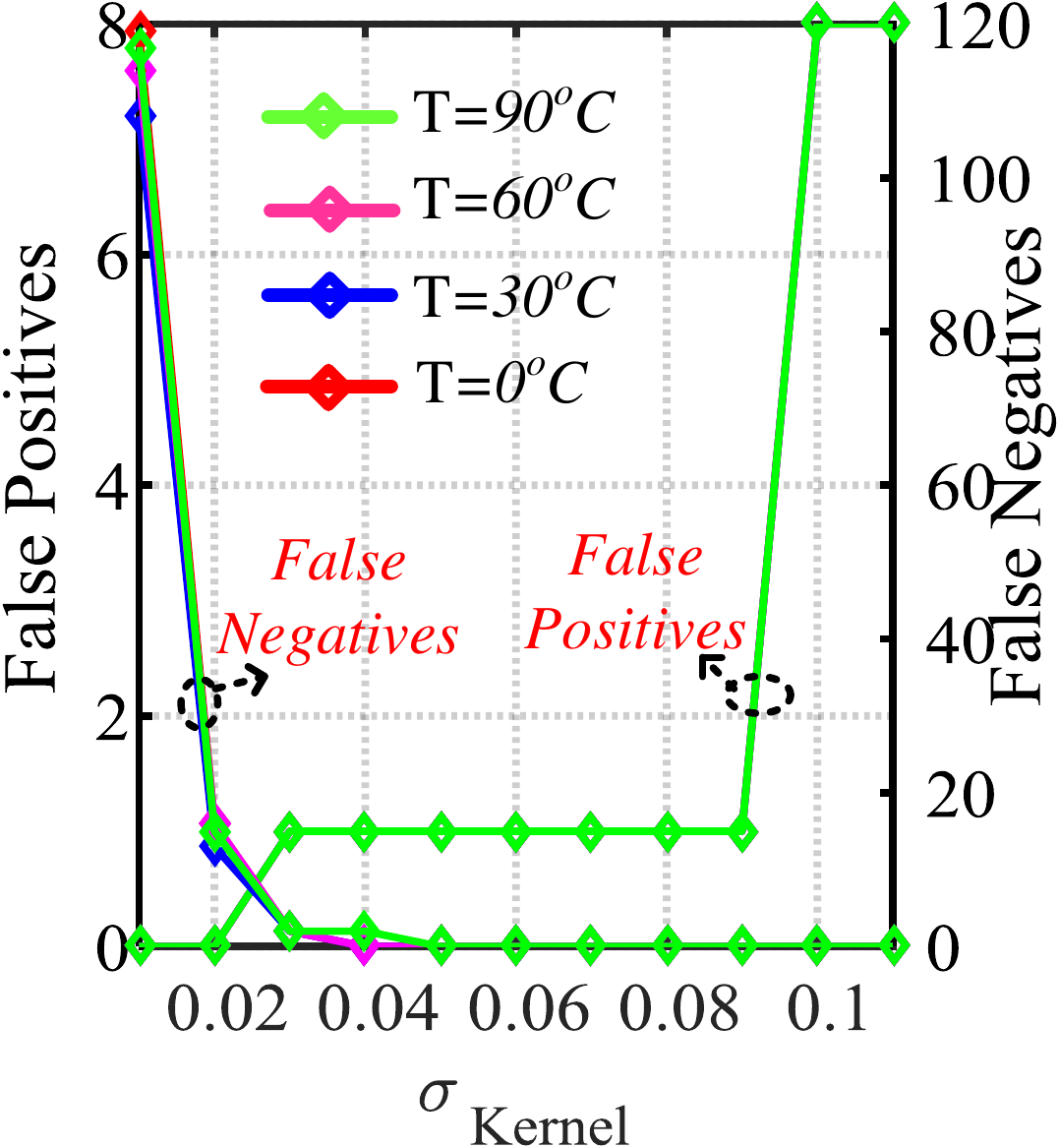}}\quad
     \caption{Impact of process and temperature induced variation on AEGIS performance.}
     \label{Accuracy_3}
\end{figure}

Further, process variation has a significant impact on the reliability of the learned PDF. Due to process variation, parameters of the Gaussian kernels such as mean ($\mu_{Kernel}$), standard deviation ($\sigma_{Kernel}$) and amplitude vary significantly. Variation in aforementioned kernel parameters due to process variability degrades the likelihood of the inlier samples while improving that of the outliers. To study the impact of process variation on AEGIS performance, we obtain the variation statistics of  Gaussian kernel for different $\sigma_{Vth}$ using HSPICE-based Monte Carlo simulations. Next, obtained distributions were used to emulate the process variation and alter corresponding parameters in algorithm-based outlier detection setup using MATLAB. In Fig. \ref{Accuracy_3}(a), $F_{N}$ and $F_{P}$ performances degrade with increasing $\sigma_{Vth}$ due to imprecise hyper-plane. AEGIS allows programming of $\sigma_{Kernel}$ to attain optimal detection efficiency.  

Additionally, temperature induced variations affect mean ($\mu_{Kernel}$) and standard deviation ($\sigma_{Kernel}$) deviation of the Gaussian kernels [Sec. V]. We have characterized the variation in $\mu_{Kernel}$ and $\sigma_{Kernel}$ at different temperatures using HSPICE-based Monte Carlo simulations. Obtained characteristics are used in MATLAB-based AEGIS setup for performance benchmarking. At lower $\sigma_{Kernel}$, deviation in $\mu_{Kernel}$ and $\sigma_{Kernel}$ will degrade $F_{N}$ due to inaccuracies in the likelihood of the inlier samples. On the other hand, at higher $\sigma_{Kernel}$, deviation in $\mu_{Kernel}$ and $\sigma_{Kernel}$ is relatively insignificant to affect the classification hyperplane. Therefore, $F_{P}$ is fairly insensitive to temperature variations [Fig. \ref{Accuracy_3}(b)]. Thus, the impact of temperature variations on AEGIS's performance can be adaptively alleviated by programming $\sigma_{Kernel}$. 

\begin{table}[ht]
\centering
\caption{f1-score performance of the discussed architecture for different time-series data with $\sigma_{Kernel}$=0.05, $\sigma_{Vth}$=15mV, $\sigma_{Noise}$=25mV, and T=90$^o$C.}
\scalebox{0.75}{
\begin{tabular}{|c|c|c|c|c|c|}
\hline
\multirow{2}{*}{Time Series} & \multicolumn{5}{c|}{f1-score}                                                \\ \cline{2-6} 
                             & P$_{Thres}$=10$\mu$ & P$_{Thres}$=100$\mu$ & P$_{Thres}$=1m & P$_{Thres}$=10m & P$_{Thres}$=100m \\ \hline
1                            & \textbf{\textcolor{black}{0.9412}}            & \textbf{\textcolor{black}{0.9412}}               & \textbf{\textcolor{black}{0.9412}}         & 0.64        & 0.26         \\ \hline
2                            & \textbf{\textcolor{black}{0.9091}}            & \textbf{\textcolor{black}{0.9091}}              & \textbf{\textcolor{black}{0.9091}}            & 0.77        & 0.63         \\ \hline
3                            & 0.67            & 0.67             & 0.67       & \textbf{\textcolor{black}{0.93}}        &  0.72           \\ \hline
4                            & \textbf{\textcolor{black}{1}}             & \textbf{\textcolor{black}{1}}              & 0.47      & 0.096        & 0.03         \\ \hline
5                            & 0.70            & 0.72             & 0.72       & 0.76        & \textbf{\textcolor{black}{0.87}}            \\ \hline
\end{tabular}}
\end{table}

\begin{table}[ht]
\centering
\caption{f1-score performance of the discussed architecture for different time-series data with $P_{Thres}$=100$\mu$, $\sigma_{Vth}$=15mV, $\sigma_{Noise}$=25mV, and T=90$^o$C.}
\scalebox{0.71}{
\begin{tabular}{|c|c|c|c|c|c|}
\hline
\multirow{2}{*}{Time Series} & \multicolumn{5}{c|}{f1-score}                                                \\ \cline{2-6} 
                             & $\sigma_{Kernel}$=0.02 & $\sigma_{Kernel}$=0.04 & $\sigma_{Kernel}$=0.06 & $\sigma_{Kernel}$=0.08 & $\sigma_{Kernel}$=0.1 \\ \hline
1                            & 0.27            & \textbf{\textcolor{black}{0.9412}}               & \textbf{\textcolor{black}{0.9412}}         & 0.7       & 0.64         \\ \hline
2                            & 0.62           & \textbf{\textcolor{black}{0.9091}}              & 0.83           & 0.83        & 0.76        \\ \hline
3                            & 0.43            & 0.67             & 0.67       & 0.67       &  0.67           \\ \hline
4                            & 0.056             & 0.4651              & \textbf{\textcolor{black}{1}}      & \textbf{\textcolor{black}{1}}       & 0.67         \\ \hline
5                            & 0.76            & 0.73             & 0.69       & 0.67        & 0.66           \\ \hline
\end{tabular}}
\end{table}

Table IV and Table V summarizes the performance of our approach with $\sigma_{Vth}$=15mV, $\sigma_{Noise}$=25mV, T=90$^o$C. From Table IV and V,
$\sigma_{Kernel}$ and P$_{Thres}$ should be co-optimized to achieve high detection efficiency and enhance the resilience of the statistical model against ambient noise, process \& temperature induced variations.

\subsection{AEGIS Power Consumption Analysis}

\begin{figure}[!t]
     \centering
\subfloat[][]{\includegraphics[width=4cm, height=4.5cm]{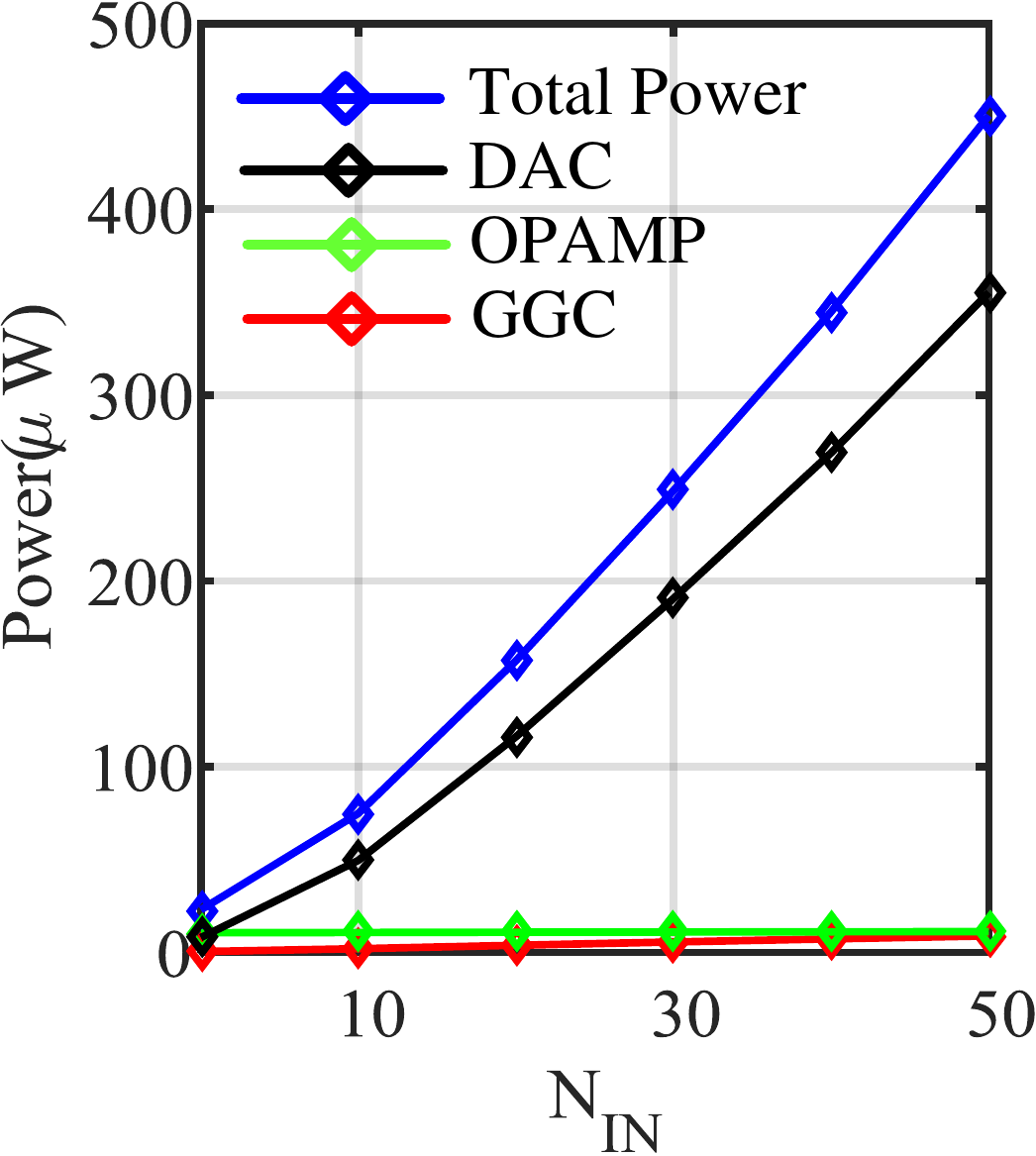}}\quad
\subfloat[][]{\includegraphics[width=4cm]{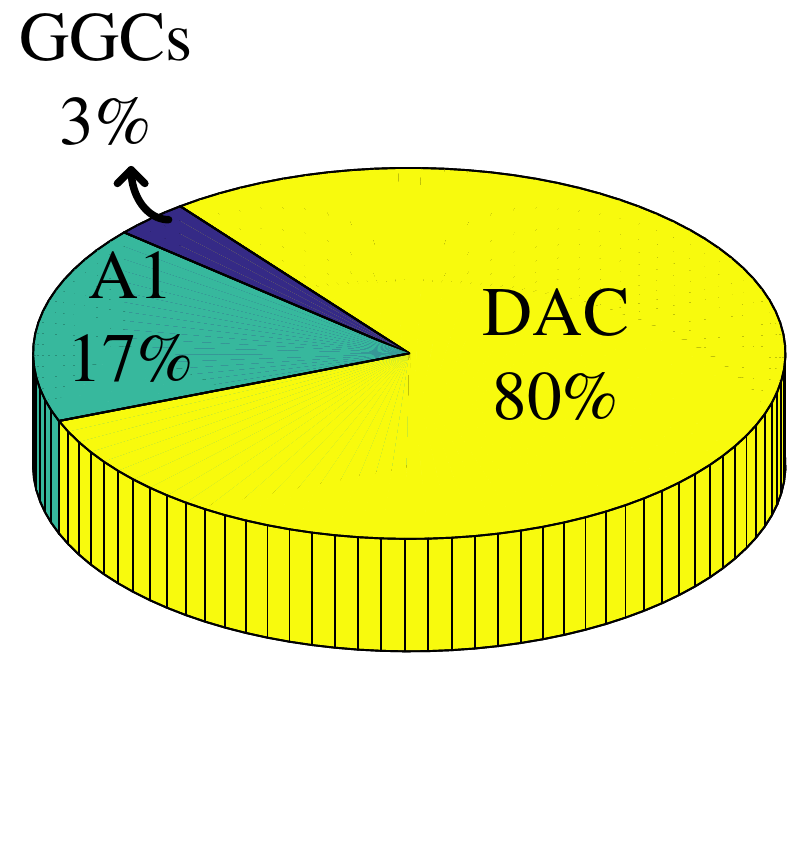}}\quad
     \caption{(a) Power consumption of different modules in AEGIS for varying sliding window length. (b) Module-wise power consumption breakdown of AEGIS architecture for N$_{IN}$=10. CSDAC array dominates the overall power consumption.}
     \label{Power}
\end{figure}

Fig. \ref{Power}(a) shows the shows the average power consumption of different modules in AEGIS for different N$_{IN}$. Power consumption of the gain-stage A1 in PDF learner is fairly insensitive to N$_{IN}$ and on an average consumes $\sim$13$\mu$W. Although power consumption of GGC array increases with N$_{IN}$, GGCs are operated in the subthreshold regime to minimize the overheads. Since the currents steering DACs are implemented using binary weighted PMOS current sources, CSDACs incur significant power overhead and the energy efficiency of the DAC array degrades with N$_{IN}$. Fig. \ref{Power}(b) shows the power consumption breakdown of different modules in AEGIS for N$_{IN}$=10. In Fig. \ref{Power}(b), CSDAC array on an average consumes 60$\mu$W while GGC array operates with $\sim$2$\mu$W. For N$_{IN}$=10, AEGIS on an average consumes $\sim$75$\mu$W while operating at a sampling frequency of 2MHz.

\begin{figure}[!t]
     \centering
     \includegraphics[width=\linewidth]{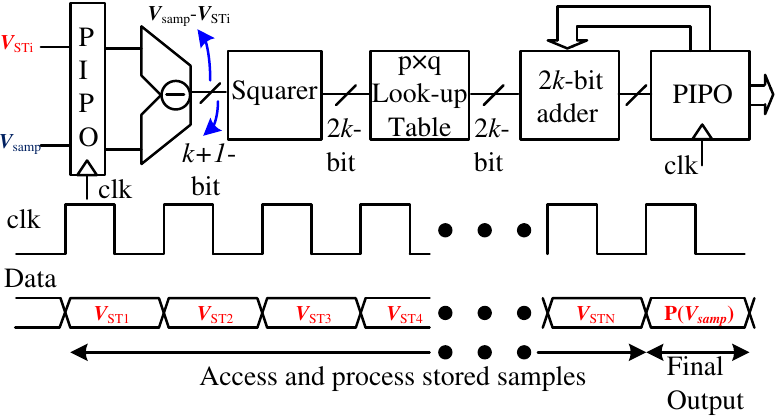}
     \caption{Digital implementation of KDE is relatively resource expensive as it requires LUT to store Exponential function samples, $2k$-bit adder, squarer circuit, $k$-bit subtractor and parallel-in-parallel-out (PIPO) registers. (Inset: \textbf{\textit{V}$_{STi}$} \& \textbf{\textit{V}$_{samp}$} represents digitized V$_{STi}$ \& V$_{samp}$, respectively)}
     \label{Digital_KDE}
\end{figure}

Further, Fig. \ref{Digital_KDE} shows the comparable digital implementation of KDE. We have synthesized the simplified architecture in Fig. \ref{Digital_KDE} (without the look-up table) using 45nm standard cells library in \cite{NCSU}  considering $k$=4. The simplified digital KDE architecture performs subtraction, squaring and accumulation in each clock cycle. RTL synthesis is performed on Cadence RC compiler to estimate worst-case power consumption at a clock frequency of 20MHz. The worst-case power consumption of the simplified digital KDE is 40$\mu$W per clock cycle. Hence, worst-case power consumption to compute likelihood of \textbf{\textit{V}}$_{samp}$ using N known RV samples can be extrapolated as 40$\mu$W$\times$N. It should be noted that 40$\mu$W$\times$N does not include look-up table (LUT) overheads. Thus, discussed mixed-signal approach is more than 4$\times$ power efficient than digital implementation

\section{Use-cases of AEGIS}
Low power outlier detection using AEGIS can find applications in a variety of IoT domains such as intrusion detection, medical sensor diagnostics, and industrial system diagnosis. AEGIS-based anomaly detection can be utilized for fault detection and diagnosis in heating, ventilation, and air conditioning (HVAC) systems. Unlike prior works \cite{munir2017pattern,srinivasan2015bugs}, AEGIS-based outlier detection can be implemented within the edge nodes. Likewise, in environment sensing and control applications, sensor anomalies can arise due to a variety of environmental artifacts \cite{derrible2018approach}. Typically such sensor anomalies are ignored, and the top-level classification layer is designed for a robust prediction. AEGIS can assist by enabling a distributed anomaly filtering to relieve the complexity of the classification layer. In medical diagnosis, readings from portable electrocardiogram (ECG) devices are noisy due to motion artifacts. Moreover, it is challenging to gather data for different abnormal heart conditions \cite{ECG_1}. Nonparametric density function modeling of AEGIS learns locally from the inlier samples and with a sufficiently large $N_{IN}$, the robustness of outlier detection can be extended to complex signal statistics. 
While the accuracy of AEGIS-based outlier detection is limited by its low complexity implementation, the design parameters such as $\sigma_{Kernel}$ and $P_{Thres}$ can be optimized to selectively suppress false negatives at the cost of more false positives [Fig. \ref{Accuarcy_1}(b)]. 


\section{Conclusion}
This work has presented novel low-power CMOS architecture for real-time ``on-sensor" outlier detection in a sensor data stream based on non-parametric statistical density estimation using Kernel Density Estimation (KDE). In our approach, Gaussian kernels for KDE-based probability density estimation are realized using low-power CMOS Gilbert Gaussian circuit (GGC) with PMOS input stage; designed at 45nm technology node. GGCs operate with a minimal power consumption of 200 nW. Also, we adopted a sliding window to update the detection model in run-time and minimize resource overhead. Yahoo database consisting of real-time series data was considered to benchmark the performance of our approach. f1-score higher than 0.87 was achieved at 75$\mu$W, operating at 2MHz sampling frequency and using only ten recent samples for density estimation, i.e., N$_{IN}$=10. Equivalent digital implementation will consume $\sim$400$\mu$W at matching throughput while neglecting the LUT overheads. Therefore, discussed mixed-signal anomaly detection framework is more than 4$\times$ power efficient than digital KDE implementation. We also showed that AEGIS allows enhancing the generalizability of the detection model depending on application specifics by programming N$_{IN}$, $\sigma_{Kernel}$, and $P_{Thres}$. 

  \balance
  \bibliographystyle{IEEEtran}
  \bibliography{main.bib}

\end{document}